\documentclass{aa} 

\usepackage{graphicx}
\usepackage{txfonts}
\usepackage{hyperref}
\hypersetup{
 colorlinks=true,
 urlcolor=blue,
 linkcolor=blue,
 citecolor=blue
}
\usepackage{adjustbox}
\usepackage{blindtext}
\usepackage{upgreek}
\usepackage{mathrsfs}
\usepackage{float}
\usepackage{xcolor}
\usepackage{url}
\usepackage{natbib}
\usepackage{textcomp}

\newcommand\CaII{\ion{Ca}{II}}
\newcommand\CaIIK{\CaII~K}
\newcommand\CaIIH{\CaII~H}
\newcommand\CaIIHK{\CaII~H\&K}
\newcommand\CaIIIR{\CaII~8542~\AA}
\newcommand\MgII{\ion{Mg}{II}}
\newcommand\MgIIK{\MgII~k}
\newcommand\MgIIhk{\MgII~h\&k}

\newcommand\FeI{\ion{Fe}{I}}
\newcommand\Halpha{{H$\alpha$}}
\newcommand\logt{{\log(\tau_{500\,\mathrm{nm}})}}
\newcommand\K{{K$_2$}}

\newcommand\kms{\mbox{km s$^{-1}$}}

\usepackage{color}
\definecolor{deepmagenta}{rgb}{0.8, 0.0, 0.8}

\begin{document} 

  \title{Physical properties of bright \CaIIK\ fibrils in the solar chromosphere}


  \author{Sepideh Kianfar\inst{1}   
  \and
  Jorrit Leenaarts\inst{1}
  \and
  Sanja Danilovic\inst{1}
  \and
  Jaime de la Cruz Rodr\'{\i}guez\inst{1}
  \and
  Carlos Jos\'{e} D\'{\i}az Baso\inst{1}
}

  \institute{Institute for Solar Physics, Department of Astronomy, Stockholm University, Albanova University Centre, 10691 Stockholm, Sweden\\
       \email{sepideh.kianfar@astro.su.se}
       }

  \date{Received \textit{month dd, yyyy}; accepted \textit{month dd, yyyy}}

 
 \abstract
  {Broad-band images of the solar chromosphere in the \CaIIHK\ line cores around active regions are covered with fine bright elongated structures called bright fibrils. The mechanisms that form these structures and cause them to appear bright are still unknown.}
  {We aim to investigate the physical properties, such as temperature, line--of--sight velocity, and microturbulence, in the atmosphere that produces bright fibrils and to compare those to the properties of their surrounding atmosphere.}
  {We used simultaneous observations of a plage region in \FeI~6301-2~\AA, \CaIIIR, \CaIIK, and \Halpha\ acquired by the CRISP and CHROMIS instruments on the Swedish 1-m Solar Telescope. We manually selected a sample of 282 \CaIIK\ bright fibrils. We compared the appearance of the fibrils in our sample to the \CaIIIR\ and \Halpha\ data. We performed non-local thermodynamic equilibrium (non-LTE) inversions using the inversion code STiC on the \FeI~6301-2~\AA, \CaIIIR, and \CaIIK\ lines to infer the physical properties of the atmosphere.}
  {The line profiles in bright fibrils have a higher intensity in their \K\ peaks compared to profiles formed in the surrounding atmosphere. The inversion results show that the atmosphere in fibrils is on average $100-200$~K hotter at an optical depth $\logt = -4.3$ compared to their surroundings. The line-of-sight velocity at chromospheric heights in the fibrils does not show any preference towards upflows or downflows. The microturbulence in the fibrils is on average 0.5~\kms\ higher compared to their surroundings. Our results suggest that the fibrils have a limited extent in height, and they should be viewed as hot threads pervading the chromosphere.
  }
 {}


  \maketitle
%
\section{Introduction}
\label{sec:int}


Observational studies of the solar chromosphere have proven that this layer is host to a wide range of fine dynamic features \citep[e.g.][]{1974soch.book.....B, 1998assu.book.....Z, judge2006,carlsson2019}. Bright fibrils seen in the \CaIIHK\ lines are highly dynamic chromospheric structures with lifetimes of about 5~minutes \citep{Pietarila09, gafeira17}. They form a carpet of jet-like stripes that are tightly packed on the solar surface. They were reported for the first time in the narrow-band \CaIIK\ observations in \citet{zirin1974} appearing together with dark fibrils in plage regions. Since then, bright fibrils in \CaIIHK\ have mostly been observed with wide-band filters where they were not spectrally resolved and hence assumed to be low-opacity features \citep{rutten06}. These structures have been referred to with different names in the literature, such as mottles or straws, depending on where they are located on the disc \citep[and references therein]{Pietarila09}. Similar bright and elongated structures have also been observed in \CaIIIR\ \citep[e.g.][]{Pietarila11} and the \Halpha\ line \citep[e.g.][]{2011ApJ...742..119R}. Bright \CaIIK\ fibrils were reported to be the same structures as bright spicules in \Halpha\ \citep{zirin1974}. 

Fibrillar structures in both \CaIIK\ and \CaIIIR\ emanate from the magnetic field concentrations in the plage regions \citep{reardon2008,Pietarila09}. Bright fibrils seen in wide-band \CaIIH\ images follow the magnetic field lines at low-chromospheric heights \citep{jafarzadeh17}. \CaIIIR\ fibrils have the same behaviour \citep{jaime11, Asensio2017}. The bright \CaIIHK\ fibrils appear thin (less than 200~km width) and are elongated with lengths up to 4000~km in the internetwork areas where they are longer and more stable \citep{Pietarila09, gafeira17} compared to the regions closer to the magnetic concentration, where they are more dynamic and shorter. Bright fibrils observed in \CaIIHK\ as well as \CaIIIR\ show different types of wave motions \citep{Pietarila11,gafeira17_2, jafarzadeh17_2}.
In previous studies, it has been suggested that bright fibrils are bright due to temperature enhancement at their location \citep[e.g. in][who used \Halpha\ line-width as a temperature proxy]{molnar2019}. 

During the past years it has become possible to perform non-local thermodynamic equilibrium (non-LTE) inversion of chromospheric spectral lines \citep{2015A&A...577A...7S,jaime16,2018A&A...617A..24M,jaime19}. These inversions aim to produce spatially resolved model atmospheres which are consistent with observations. They provide a direct physics-based, as opposed to proxy-based, way to determine temperatures in the chromosphere.

In this paper we use high spatial and spectral resolution data in \CaIIK\ observed simultaneously with \Halpha\ and full-polarimetric data taken in \FeI, \CaIIIR\ lines taken with the Swedish 1-m Solar Telescope. This dataset allows us to study the spectral profiles of the \CaIIK\ bright fibrils and to compare them with other chromospheric lines. In addition, we apply non-LTE inversions to the dataset in order to retrieve the physical parameters, such as temperature and line-of-sight velocity of a sample of bright fibrils. 

\section{Observations }
\label{sec:obs}

\begin{figure*}
  \centering
  \includegraphics[width=\linewidth]{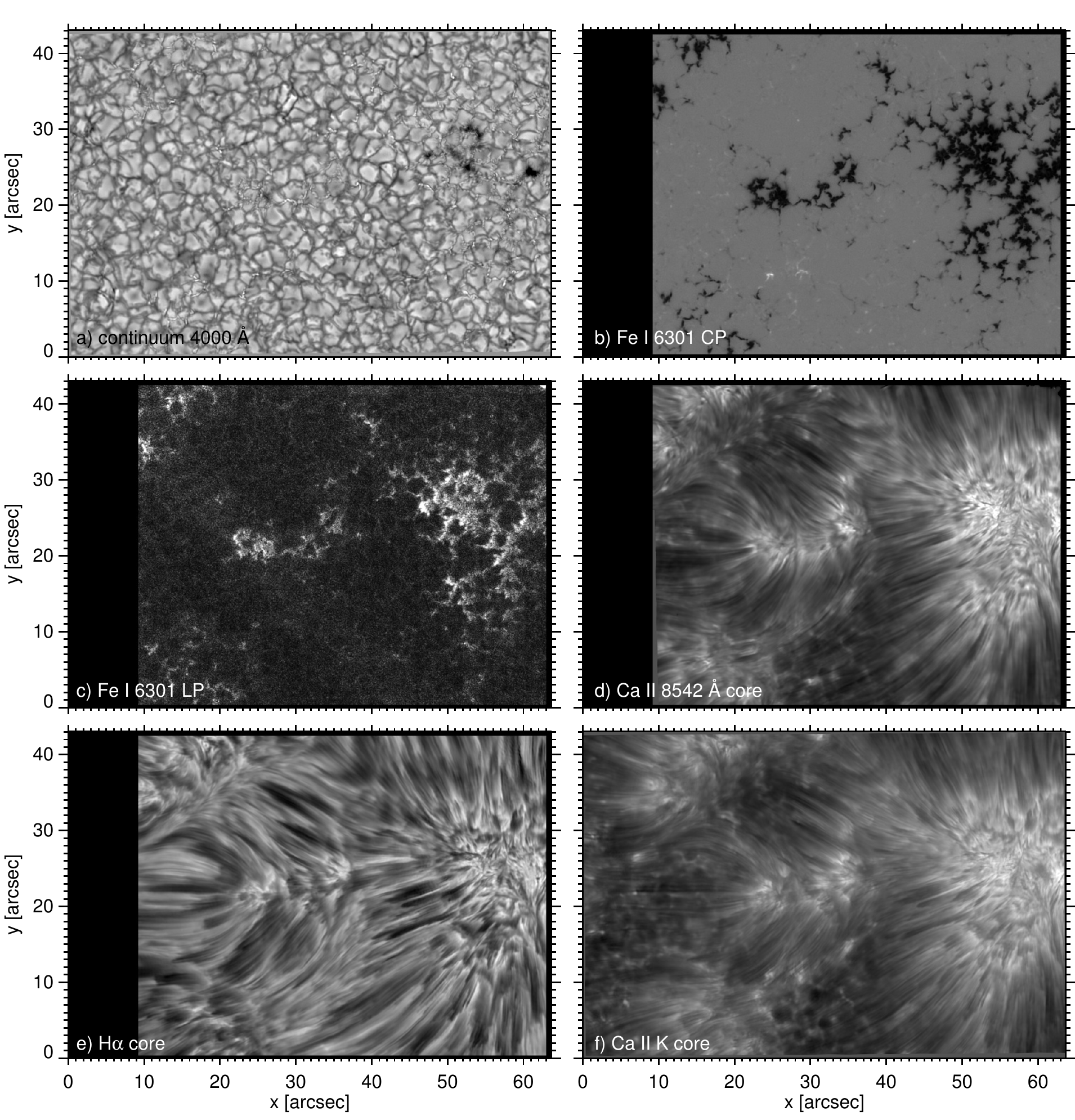}
  \caption{Overview of the observations taken on 2016-09-15 at 08:57:02~UT. \textit{a)} continuum intensity at 4000~\AA; \textit{b)} wavelength-averaged circular polarisation map based on the Stokes~$V$ signal in the \FeI~6301 line; \textit{c)} wavelength-averaged linear polarisation map based on the Stokes~$Q$\&$U$ signal in the \FeI~6301 line; \textit{d)} \CaII~8542~\AA\ nominal line centre intensity; \textit{e)} \Halpha\ nominal line centre intensity; \textit{f)} \CaII~K nominal line centre intensity. Panels d and f have been gamma-corrected to increase visibility of the structures present in the images.}
	\label{fig:FOV}
  \end{figure*}

\subsection{Data reduction}
\label{sb:data}

We used observations taken with the Swedish 1-m Solar telescope
\citep[SST;][]{scharmer03}
on 2016-09-15, at 08:49:51--09:07:39~UT. The observations were obtained with the CRisp Imaging Spectro-Polarimeter \citep[CRISP;][]{crisp} and CHROMospheric Imaging Spectrometer \citep[CHROMIS;][]{2017psio.confE..85S} instruments simultaneously. The observed field of view (FoV) of CHROMIS covers an area of $63 \arcsec \times 42 \arcsec$ which overlaps with the CRISP FoV. It contains a plage region near disc centre at helioprojective-Cartesian coordinates $(X,Y)=(-135\arcsec,77\arcsec)$, corresponding to $\mu=0.99$.

The CRISP instrument sampled the \FeI~6301 \& 6302~\AA\  line pair with 16 non--equidistant wavelength points spanning from $-0.18$ to $+1.08$~\AA\ from the 6301~\AA\ line centre. The \CaII~8542~\AA\ line, also observed with CRISP, was sampled with 19 wavelength positions with steps of 85~m\AA\ around line core, in addition to two points at $\pm 1.7$~\AA\ from the line centre. Observations in these lines were acquired with full-Stokes polarimetry. Furthermore, the \Halpha\ 6563~\AA\ line was observed in 16 non--equidistantly spaced wavelength points, covering a span of $-1.95$ to $+1.55$~\AA\ from the line centre. The CRISP data is sampled with a pixel scale of $0\farcs059$ and its cadence is 37~s.

The \CaIIK\ line was sampled by the CHROMIS instrument within a wavelength range gridded in 19 positions around the line centre (i.e. 3933.682~\AA), with 78~m\AA\ spacing and two wing points at $\pm 1.331$~\AA\ from the line centre. One further point in the continuum at 4000~\AA\ was also observed by CHROMIS. The CHROMIS data has a cadence of $\sim$~15~s and a pixel size of 0$\farcs$038.

The raw data were reconstructed using the CRISPRED
 \citep{crispred}
 and CHROMISRED 
\citep{chromisred}
pipelines for CRISP and CHROMIS data, respectively. The reduction process applies the image restoration method by using Multi-Object Multi-Frame Blind Deconvolution
\citep[MOMFBD;][]{momfbd}. 
Afterwards, the CRISP and CHROMIS data were co-aligned (we estimate the accuracy to be within a tenth of a pixel size). The CRISP data was then resampled at the CHROMIS pixel scale. The CRISP cadence was adjusted to CHROMIS cadence by using the nearest--neighbour interpolation. The last step in the post--process procedure was to perform absolute intensity calibration on the data using the atlas profile from 
\citet{atlas}.

\subsection{FoV synopsis}
\label{sb:FOV}

Due to the variability of the seeing during the observation, we chose to analyse only a single scan taken at 08:57:02~UT, which had good seeing conditions. Figure~\ref{fig:FOV} shows an overview of the selected scan. 
The FoV contains a unipolar plage region on the right, with a number of other patches of the same polarity elsewhere in the FoV. The plage contains some small pores and many of the field concentrations show up as facular brightenings in the intergranular lanes in the continuum. There are a few small-scale field concentration of opposite polarity located away from the plage.

Figure~\ref{fig:FOV}d, e and f show a carpet of dark and bright fibrils emanating from the plage that cover the FoV in the \CaIIIR, \Halpha\ and \CaIIK\ line cores. They appear more prominent and optically thick in the \Halpha\ panel than in the \CaII\ panels. Part of this difference is caused by the difference in line width: pixel-to-pixel variations in vertical velocity Doppler-shift the core of the \CaII\ lines more than the core of \Halpha, owing to the larger atomic mass of Ca. Therefore, the different appearance of the structures in different lines is intrinsic: the fibril layer has a larger optical thickness in \Halpha\ than in \CaIIK\ and \CaIIIR\ at their nominal line centres.

There are many fibrils that appear bright in the line cores of the \CaII\ lines. They are more tightly packed and have higher intensity close to the magnetic concentrations compared to areas further away from the plage, and they appear optically thick there. Away from the plage, the fibrils appear less bright, and they appear more optically thin; one can for instance see bright intergranular lanes from the reversed granulation through the fibrillar structures.
In this article we focus on the characteristics of the bright fibrils in \CaIIK, but we also compare their appearance to other chromospheric lines in our data. We use the \FeI\ 6301-2 lines only for the inversions in Sect.~\ref{sb:inversion}.

\section{Results}
\label{s:results}

\subsection{Sample of bright fibrils}
\label{sb:sample}

\begin{figure*}
  \centering
  \includegraphics[width=\linewidth]{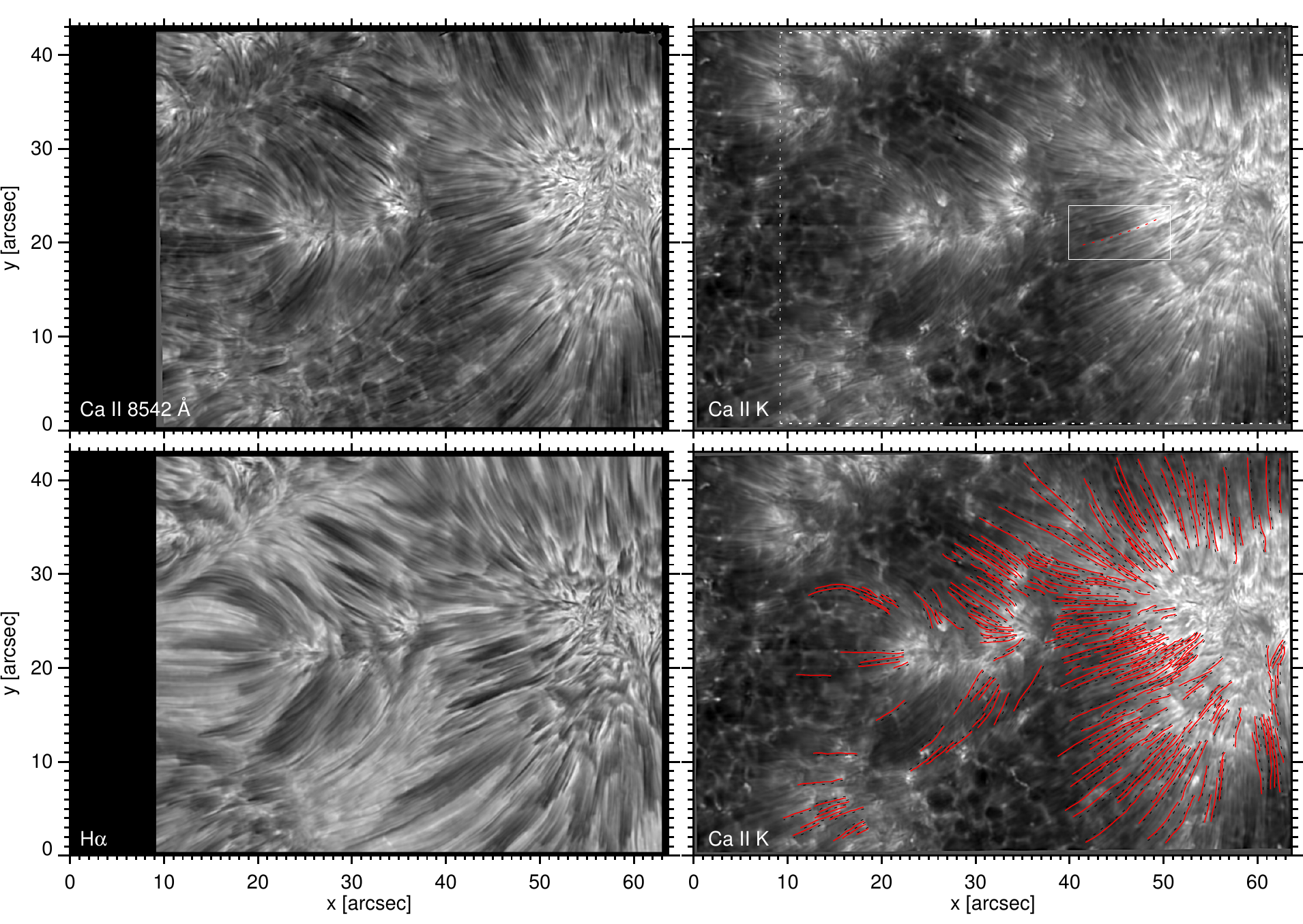}
  \caption{Wavelength-summed unsharp-masked maps of \CaIIIR, \CaIIK\ and \Halpha\ intensities; The dotted box in the \textit{top-right} panel marks the FoV of CRISP shown in Fig.~\ref{fig:FOV_inv}. The solid box shows the RoI shown in Figs.~\ref{fig:category}, \ref{fig:fb}, and \ref{fig:ROI} and the fibril indicated in it with red dotted curve is discussed in Figs.~\ref{fig:cak}, \ref{fig:rf}, and \ref{fig:fb}. The paths of all the selected fibrils in our sample are overplotted with red curves and their nearby dark lanes are shown with black dotted curves in the \textit{bottom-right} panel.}
	\label{fig:path}
  \end{figure*}

\subsubsection{Identifying the structures}
\label{sbb:indentify}

The fibrils are chromospheric structures visible mainly in near-line-core wavelengths of the chromospheric lines. Their visibility is sensitive to Doppler shifts, thus not all fibrils are visible in a map at a fixed single wavelength. Therefore, to obtain a sample of \CaIIK\ fibrils that includes as many of them in a single frame as possible, we constructed wavelength-summed maps. We summed the \CaIIK\ line profile from $-313$ to $+313$~m\AA\ around the line centre. We also summed intensity maps of the other chromospheric lines in our data for comparison: the \Halpha\ intensity over $[-450,+450]$~m\AA\ and the \CaIIIR\ intensity over $[-255,+255]$~m\AA\ around their line cores. The resulting maps have low contrast and hence to bring out the fibrillar structures, we unsharp-masked the wavelength-summed data using a filter of 10~pixels wide (i.e. slightly larger than the typical width of the bright fibrils; see Sect.~\ref{sec:int}).
The resulting maps are shown in Fig.~\ref{fig:path}.

We applied a manual approach for selecting the bright fibrils in \CaIIK\ wavelength-summed unshap-masked intensity map, using the path-defining tool in the CRisp SPectral EXplorer computer program (CRISPEX; \citealt{crispex,chromisred}). First, we identified a sample of 282 \CaIIK\ fibrils that appeared consistently bright compared to their surroundings. Then, in order to be able to investigate the physical properties of the atmosphere at the location of the bright fibrils and their nearby regions, we also specified dark lanes close to each of the fibrils in our sample referred to as \textit{background} hereafter. The selected fibrillar paths and their backgrounds are shown in the bottom-right panel of Fig.~\ref{fig:path}.

\subsubsection{Characteristics of the fibrillar \CaIIK\ line profile}
\label{sb:cak_profile}

\begin{figure*}
\centering
\includegraphics[width=0.33\linewidth]{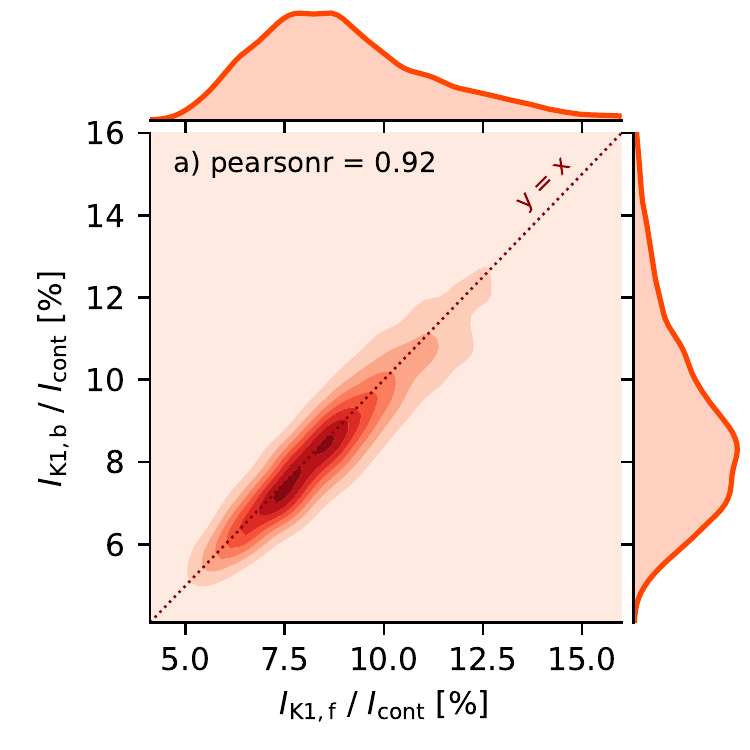}
\includegraphics[width=0.33\linewidth]{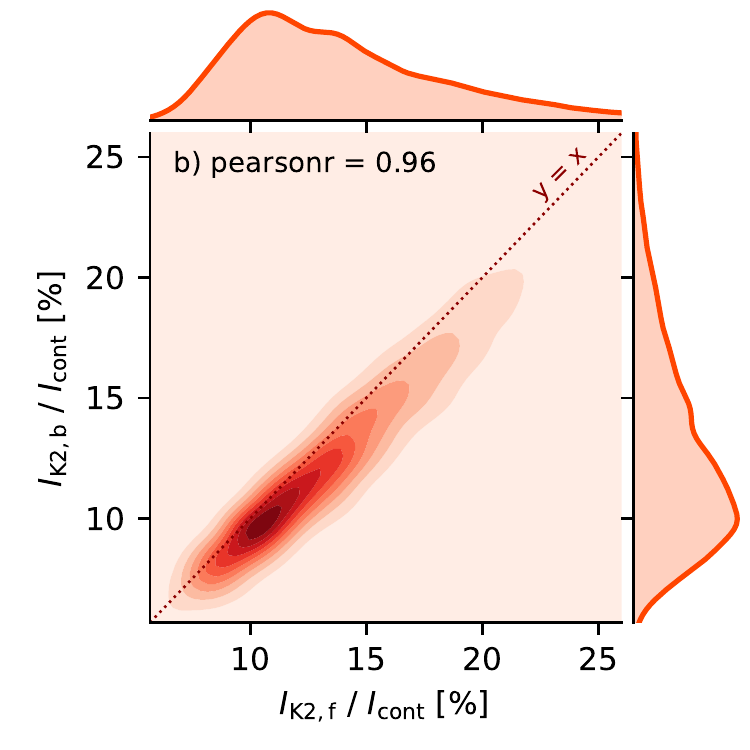}
\includegraphics[width=0.33\linewidth]{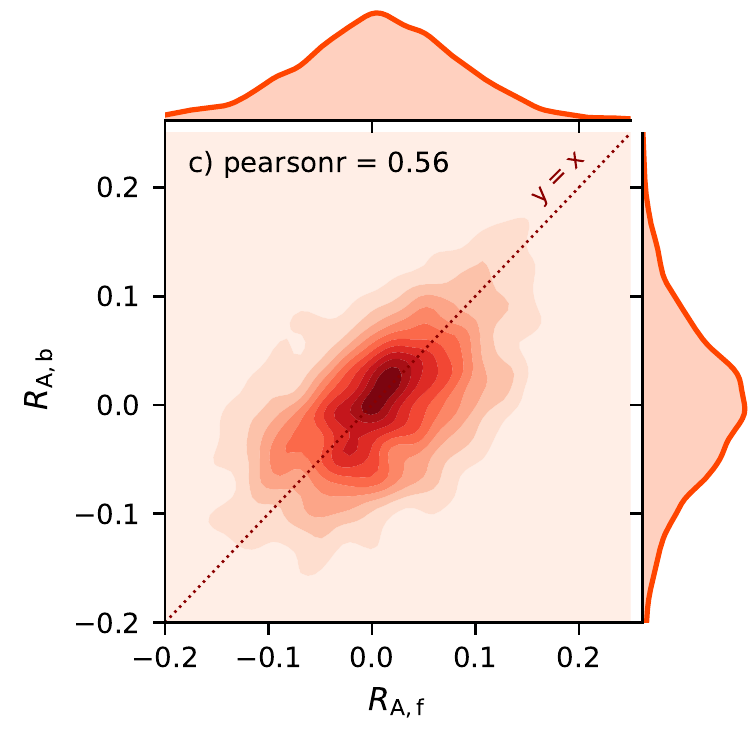}
\includegraphics[width=0.33\linewidth]{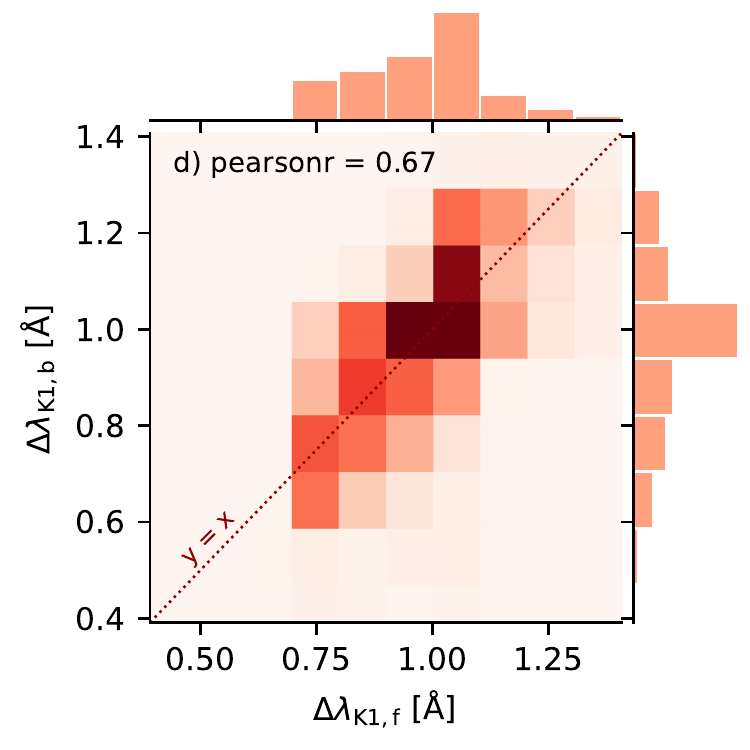}
\includegraphics[width=0.33\linewidth]{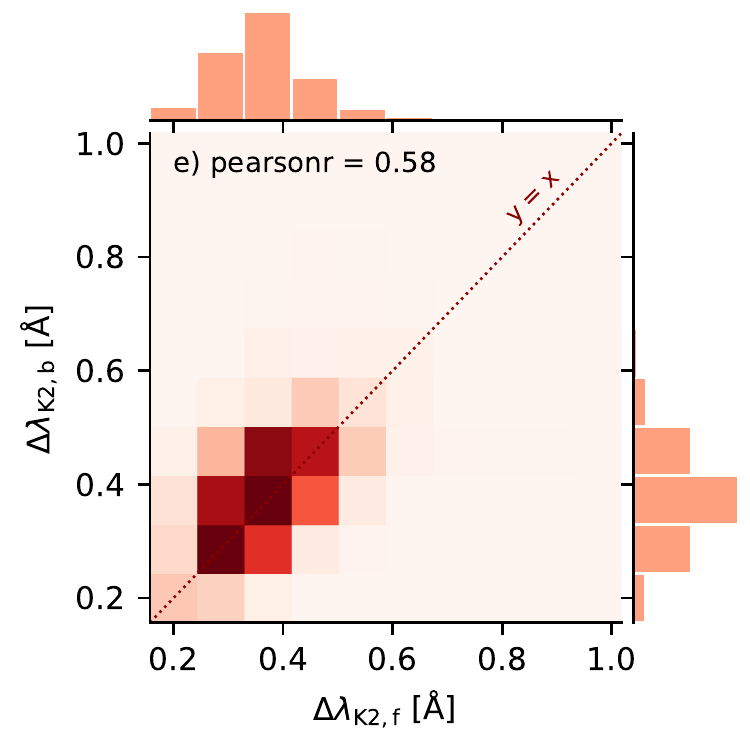}

   \caption{Joint probability distributions (JPDs) of \CaIIK\ line properties of fibrillar pixels and their background counterparts that have double-peaked profiles; panels \textit{a)} and \textit{b)} show the JPD of the average intensity of the K$\mathrm{_{1V}}$ minima and the \K\ peaks in the fibrillar pixels and their backgrounds. The average intensity value of the K$\mathrm{_{1V}}$ and \K\ features are normalised to the local continuum intensity. The distribution of the \K-peak asymmetry is shown on panel \textit{c)}. Panels \textit{d)} and \textit{e)}, show the K$_{1}$ and \K\ wavelength separations in the pixels of fibrils and their backgrounds; the bin width in these two panels is 78~m\AA\ that is, the spectral sampling size of CHROMIS around the \CaIIK\ line core. The correlation between the quantities is represented by the Pearson correlation coefficient $r$ on each panel. The $y=x$ line is overplotted for comparison.
          }
     \label{fig:k2}
\end{figure*}

\begin{figure}
  \includegraphics[width=\linewidth]{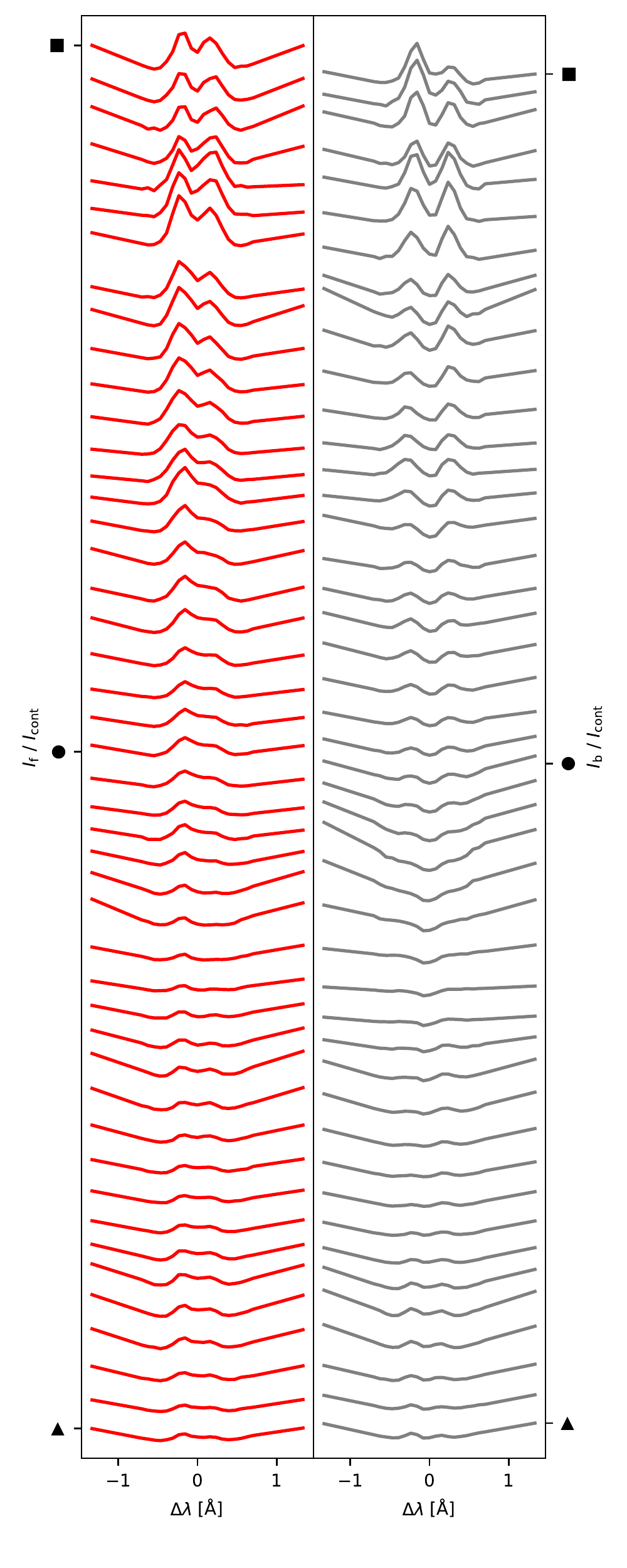}
  \caption{\CaIIK\ intensity profiles at the pixels on the fibril marked in the top-right panel of Fig.~\ref{fig:path} (\textit{left}) and the its background (\textit{right}); The square, circle and triangle symbols mark the head (close to magnetic concentration), midpoint and the tail of the structures, respectively. }
	\label{fig:cak}
  \end{figure}

To have an overview of the properties of \CaIIK\ line profiles of the fibrils (\textit{f}) compared to their backgrounds (\textit{b}), we classified all the \CaIIK\ intensity profiles using a feature finding algorithm similar to that used in \citet{jorrit2013, jorrit18} and \citet{johan18}. 
The algorithm is designed to detect the intensity and wavelength position of the \CaIIK $_{1}$ minima and \K\ emission peaks (notations adapted from \citealt{hale1904}). The results showed that 64\% of all the fibrillar pixels in our sample have profiles with two detectable \K~peaks. The rest consists of single-peaked or pure absorption profiles. In order to compare the fibrillar profiles to their surroundings, we also applied the same detection method to the pixels of the backgrounds. We found out that 82\% of the double-peaked fibrillar pixels have adjacent background pixels with double peaks in their profiles as well.

Figure.~\ref{fig:k2} shows comparisons of the quantities retrieved from these double-peaked profiles found in \CaIIK\ fibrils and their background counterparts.
The joint probability distributions in Fig.~\ref{fig:k2}a and b, show a strong correlation between the average intensity of not only the \K~peaks in fibrils and their backgrounds, but also between the intensity of K$\mathrm{_1}$ features. However, there is a small tendency towards the higher intensity in the fibrillar pixels; this tendency is more considerable in the intensity of the \K~peaks which show that the bright \CaIIK\ fibrils are mostly a line core phenomena.

The K$\mathrm{_1}$ separation varies in a range of $[0.39,1.33]$~\AA\ with an average of 1~\AA, and the \K\ separation varies from 0.16~\AA\ (i.e. the smallest $\Delta$\K\ that the algorithm is capable of detecting the peaks) to 1.02~\AA. The average \K\ separation is about 0.33~\AA\ for both fibrils and the dark lanes. The bright fibrils and their backgrounds are fairly correlated in both K$\mathrm{_1}$ and \K\ separation wavelengths and there is no systematic difference between them, as evidenced by the blob-shape of the distributions in panels d and f.

We also studied the \K-peak asymmetry calculated as follows \citep{jorrit2013}

\begin{equation}
R_A = \frac{I_{K2V} - I_{K2R}}{I_{K2V} + I_{K2R}}.
\end{equation}

A positive asymmetry ratio can be an indicator of an upflow in the lower chromosphere and a downflow in the upper layers of the chromosphere \citep{johan18}. As shown in the panel c of the Fig.~\ref{fig:k2}, the distribution of the peak asymmetry is symmetric around zero for both the fibrils and the background profiles. Again there is a fairly strong correlation between the asymmetries in the fibrils and the background, and there are no systematic differences between their asymmetries (See also Sect.~\ref{sbb:FOV}). 

To further investigate the fibrillar \CaIIK\ intensity profiles, we extracted the \CaIIK\ profile of one example fibril and its background. We studied the variation of the intensity along their paths. The fibril is marked with a red dotted curve in the top-right panel of Fig.~\ref{fig:path} and the spectral profiles are shown in Fig.~\ref{fig:cak}. The \CaIIK\ intensity profiles are double-peaked in nearly all the pixels along the fibril and in most of the dark neighbouring pixels. The fibrillar pixels, which are close to the head mark, have higher intensity than the background not only in the \K~peaks and the line-core, but also in the wings of the profile. The $\mathrm{K_{2V}}$ and $\mathrm{K_{2R}}$ are equally peaked, while in the first three profiles of the background lane $\mathrm{I_{K2V}}$ is stronger than $\mathrm{I_{K2R}}$. As we move towards the mid-point, the fibrillar pixels tend to have stronger peaks at $\mathrm{K_{2V}}$ while the scenario reverses in the background lane in which there are pixels with stronger $\mathrm{K_{2R}}$ peaks. This positive peak-asymmetry in the fibrillar pixels of this example is also confirmed by the inversion results for the line-of-sight velocity of this fibril in Sect.~\ref{sbb:fibril}. After passing the midpoints of the structures, the profile of the background turns to a pure absorption profile, while the fibril remains double-peaked with $\mathrm{I_{K2V}}>\mathrm{I_{K2R}}$ except in few pixels on the way at which the K$\mathrm{_{2R}}$ peak is either not present or too weak to be resolved given the spectral sampling size near the \CaIIK\ line-core (i.e. 78~\AA). The intensity of the fibrillar pixels decreases in the tail of the fibril and appears to be the same as the background profiles, which is by our definition the end point of the fibril, i.e. where the structure is no longer brighter than its surrounding background. The $\mathrm{K_2}$ separation of the fibrillar profiles in the first half of the bright structure is noticeably smaller compared to its background lane. This can be an indicator of lower microtubulent velocity in those fibrillar pixels \citep{solanki1991}, or a higher-located chromospheric temperature rise \citep[see Fig. 20 of][]{johan18}.

\subsubsection{Comparing the appearance of \CaIIK\ fibrils to \Halpha\ and \CaIIIR}
\label{sb:category}

\begin{figure}
  \centering
  \includegraphics[width=\hsize]{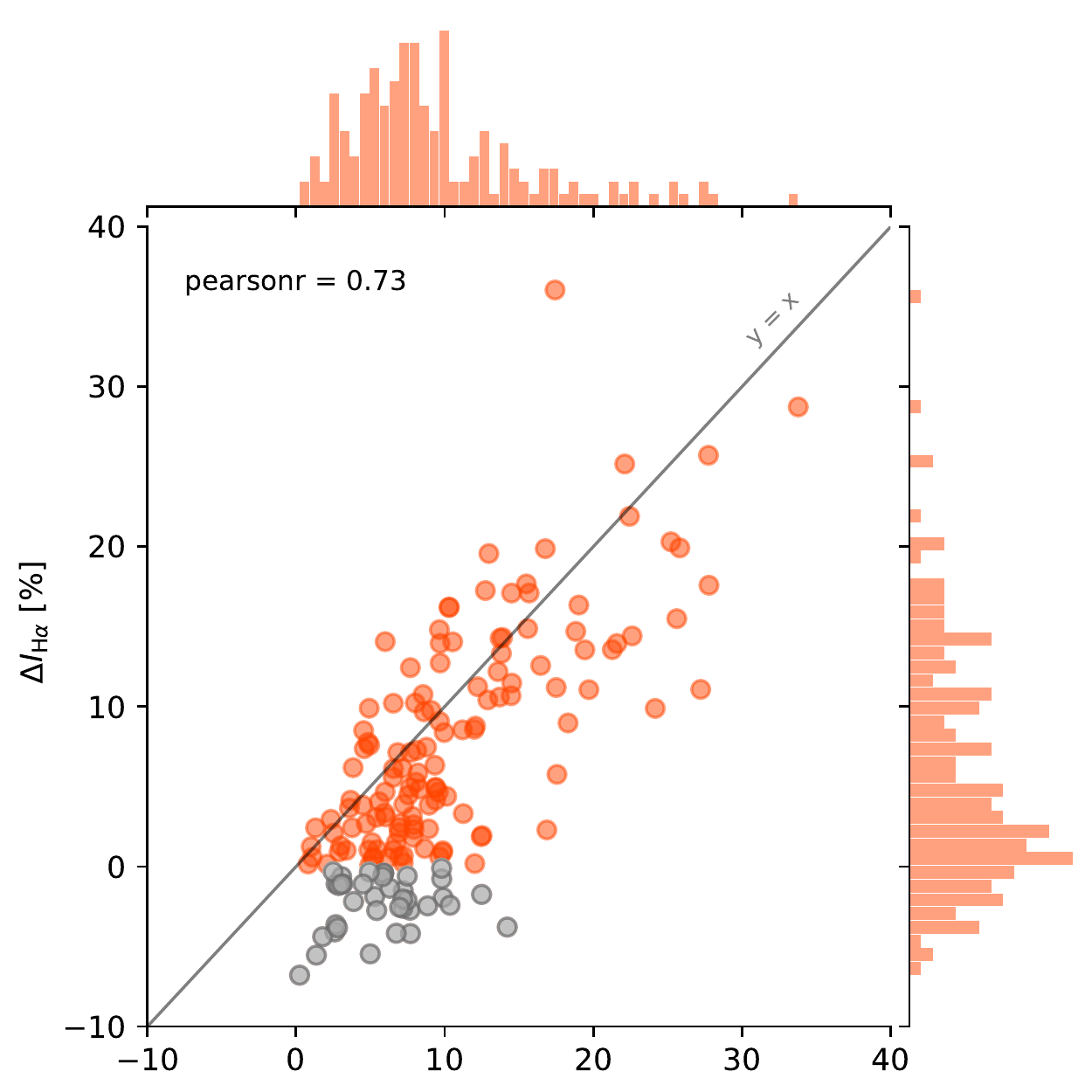}
  \includegraphics[width=\hsize]{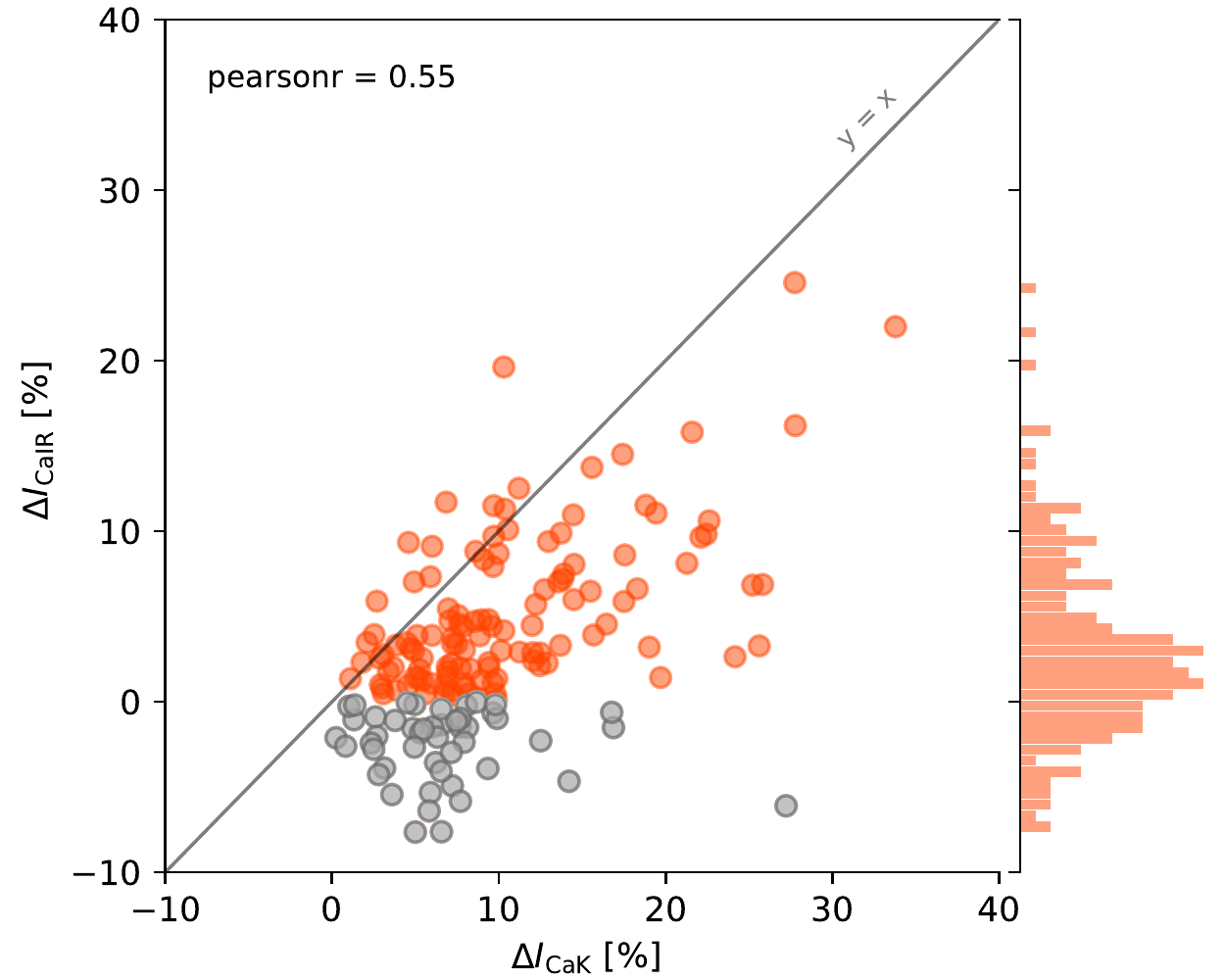}
   \caption{Comparison of the average intensity contrast of bright fibrils in \CaIIK\ versus \Halpha\ (\textit{top}) and \CaIIIR\ (\textit{bottom}); The orange-red filled circles show the fibrils which are bright in both \CaIIK\ and the other line. The grey circles show the ones that are only bright in \CaIIK. The distribution of the average intensity difference of the fibrils and their backgrounds are shown in the side-histograms. The Pearson correlation coefficient $r$ is annotated on each panel. The dashed lines show the line $y=x$.
       }
     \label{fig:I_dif}
\end{figure}

\begin{figure}
  \centering
  \includegraphics[width=\hsize]{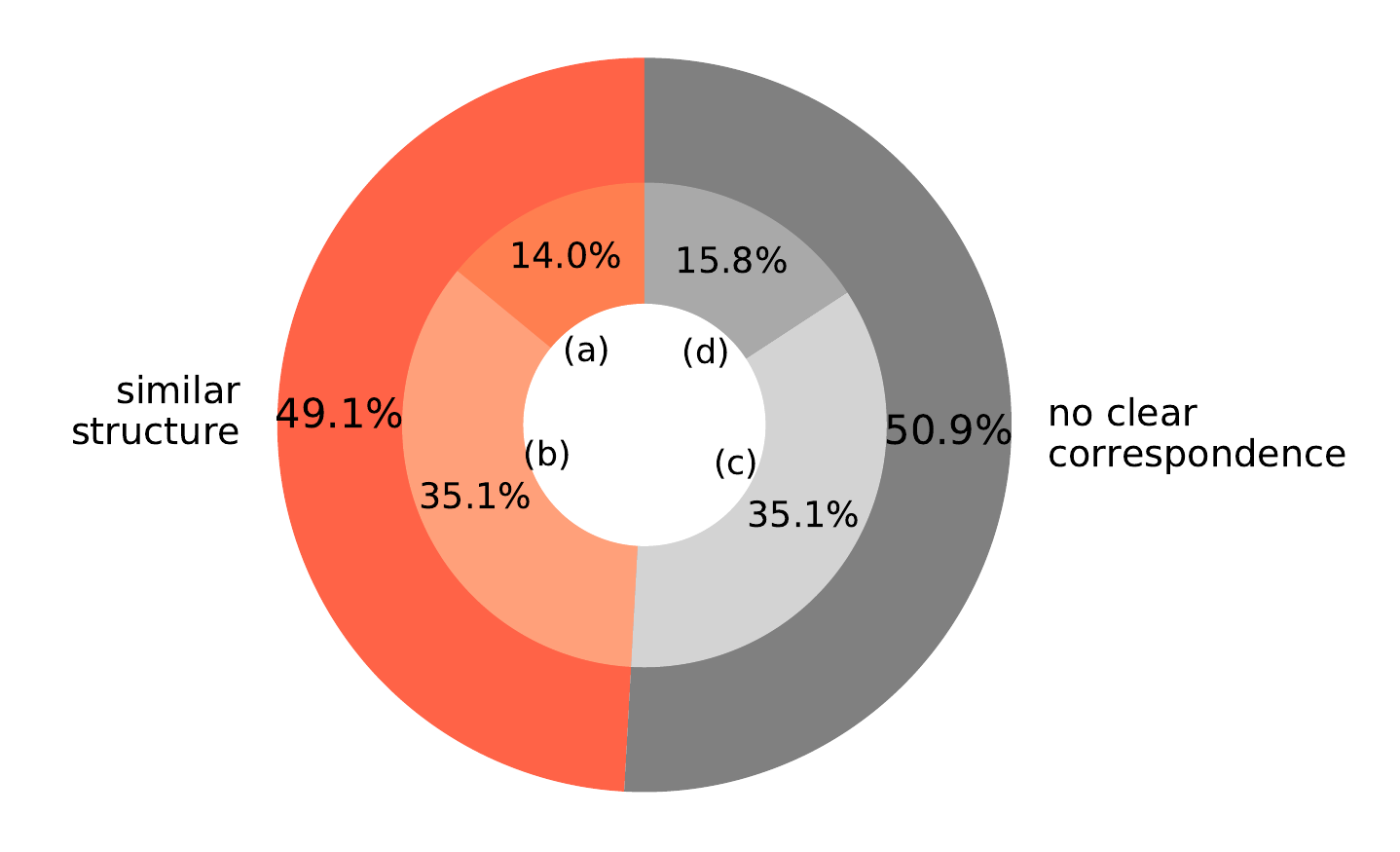}
   \caption{Categories in our fibril sample introduced in Sect.~\ref{sb:category}, presented as a nested pie plot; the inner circle shows the four categories and the outer circle groups the categories into those that have similar appearance in \CaIIK\ compared to \CaIIIR\ and \Halpha, and those that do not appear the same.
       }
     \label{fig:pie}
\end{figure}

\begin{figure*}
  \centering
  \includegraphics[width=0.385\linewidth]{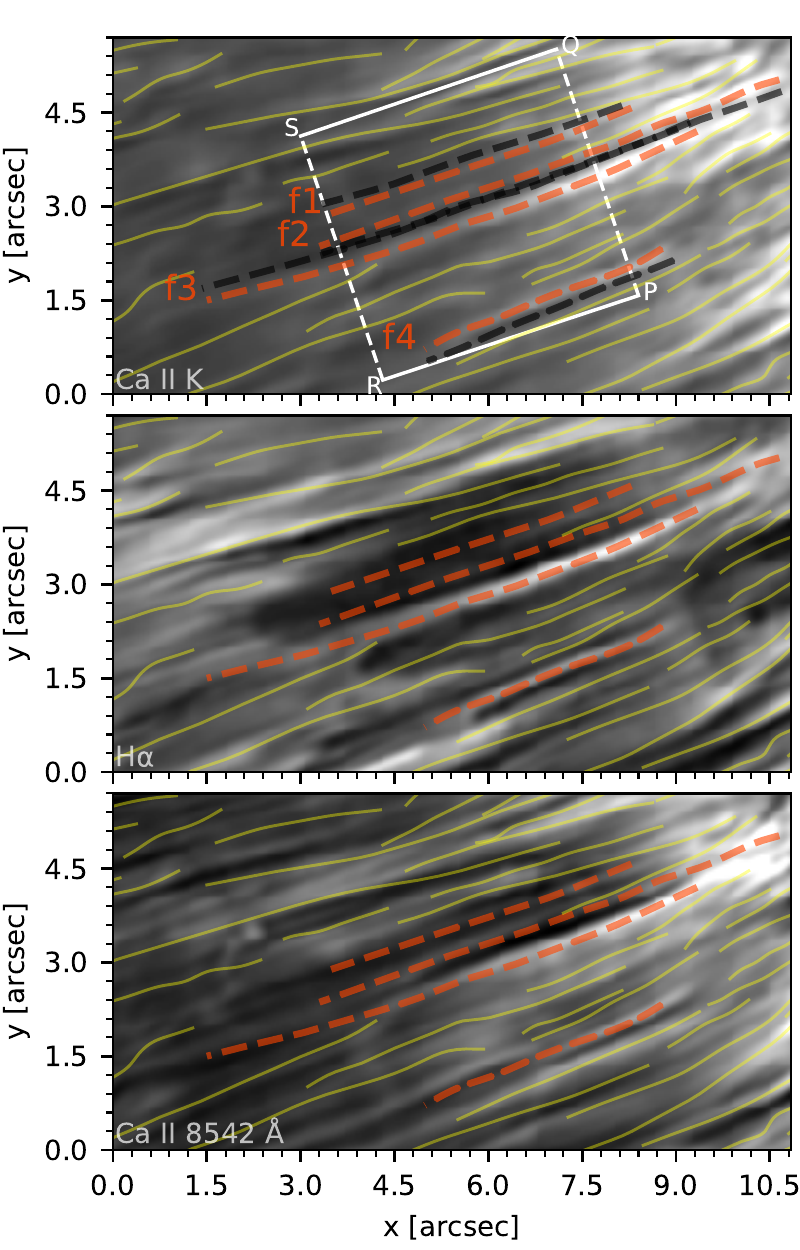}
  \includegraphics[width=0.59\linewidth]{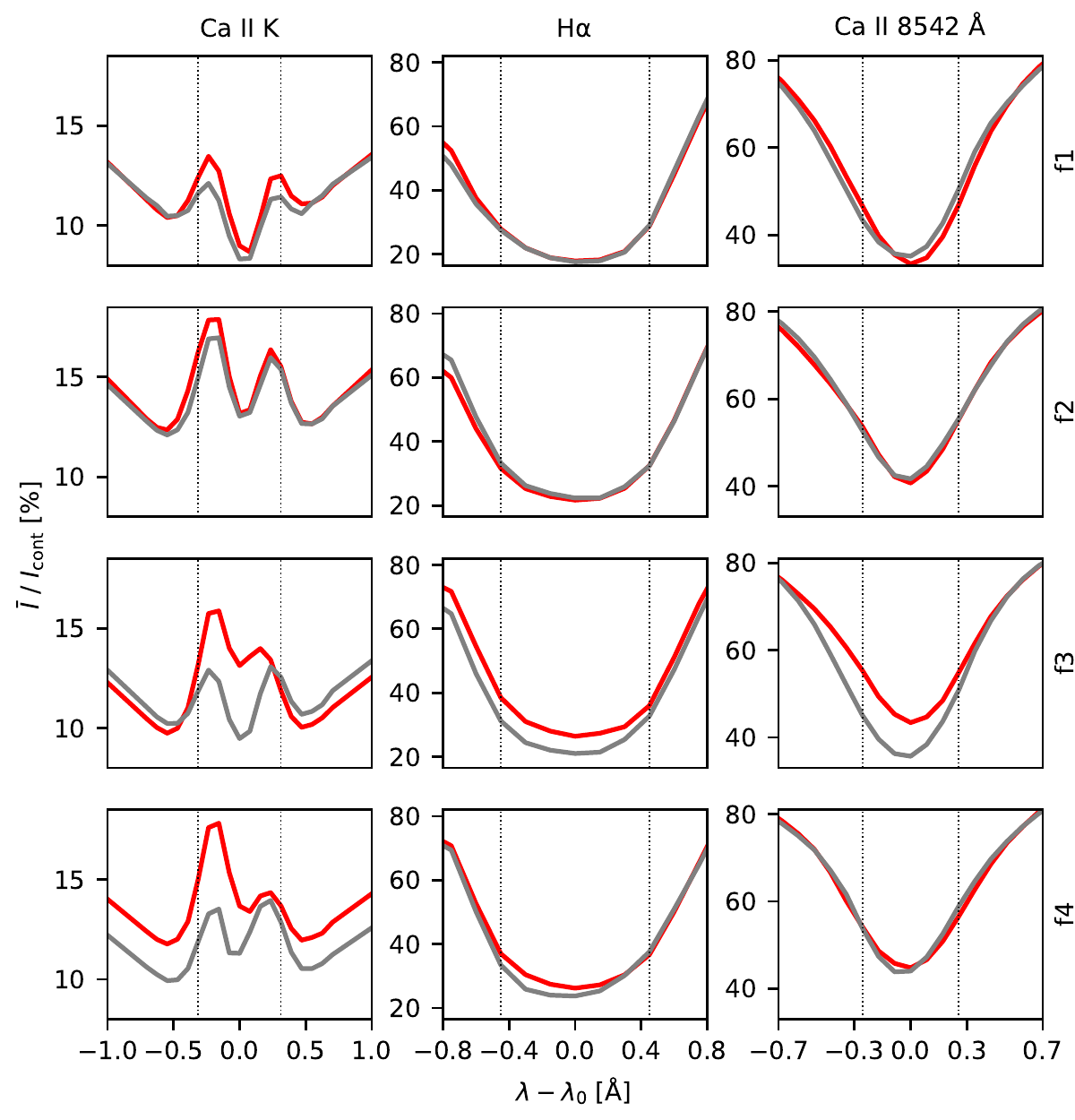}
  \caption{Examples of each of the four categories of fibrils; The \textit{left} panels show the $\lambda$-summed \CaIIK, \Halpha\ and \CaIIIR\ intensity in the RoI marked with a solid-lined box in Fig.~\ref{fig:FOV}. The example fibrils are overplotted with red-dashed lines. Their category label and nearby background lanes (grey dashed) are marked on the \CaIIK\ panel. The rest of the fibrils in our sample present in this region are indicated by yellow lines. The PQ and RS cuts are shown in Fig.~\ref{fig:across} and the PQSR box is visualised in Fig.~\ref{fig:cube}.
  The spectral profiles on the \textit{right} panels show the \CaIIK, \Halpha\ and \CaIIIR\ intensity averaged over the length of the fibril (red) and the average line profile of the corresponding dark lane (grey). 
  The dotted vertical lines mark the wavelength integration range used to produce the $\lambda$-summed intensities.}
	\label{fig:category}
  \end{figure*}

We investigated the sample to see if there are any similarities between the bright fibrils visible in the \CaIIK\ intensity map and their corresponding location in \CaIIIR\ and \Halpha\ data. First, we calculated the difference between the average intensity over the length of the \CaIIK\ bright fibrils and their background lanes ($\Delta \bar{I}_{\mathrm{CaK}}$) in the wavelength-summed map. Then we calculated the same quantity but now using the \CaIIIR\ ($\Delta \bar{I}_{\mathrm{CaIR}}$) and \Halpha\ ($\Delta \bar{I}_{\mathrm{H}\alpha}$) wavelength-summed intensities. We compared the results in Fig.~\ref{fig:I_dif}. Given that our sample includes only the fibrils that appeared brighter than their background in \CaIIK, the value of $\Delta \bar{I}_{\mathrm{CaK}}$ is always positive, but this is not the case for intensity difference in \CaIIIR\ and \Halpha: 72\% of the \CaIIK\ fibrils appear bright in \CaIIIR\ and 78\% of the \CaIIK\ bright fibrils are also bright in \Halpha. 

The above analysis is an automatic procedure solely based on the difference of the average intensities. Thus, in order to have a better comparison in terms of the appearance of the structures, we also visually compared each individual case of the \CaIIK\ fibrils in our sample to the \CaIIIR\ and \Halpha\ wavelength-summed intensity maps. 

Our investigations showed that at the location of half of the \CaIIK\ bright fibrils in our sample, there is a corresponding bright feature of similar length and orientation in other chromospheric intensity maps. In the other half of the cases, there is no clear one-to-one correspondence between the bright fibrils in different chromospheric lines of our data. We divided our sample of \CaIIK\ bright fibrils into four categories as following:

\begin{enumerate}[(a)]
  \item Same bright structure (i.e. with same length, orientation, curvature and location) as in \CaIIK\ appears in both \CaIIIR\ and \Halpha\ maps (14\%).
  \item Bright structure similar to the \CaIIK\ fibril exist in \CaIIIR\ and \Halpha\ maps but either or both of them overlap only partially with the \CaIIK\ fibril (35\%).
  \item There is a bright structure corresponding to \CaIIK\ fibril only in one of the other chromospheric lines (35\%).
  \item There is no bright structure, neither in \CaIIIR\ and nor in \Halpha\ intensity maps at the location of the bright \CaIIK\ fibril (16\%).\footnote{We note that there are cases in which the bright fibrils that exist in either of \CaIIIR\ or \Halpha\ intensity maps have no corresponding structure in \CaIIK. However, since we selected our sample based on the \CaIIK\ data only, those cases are not included in these categories.}
\end{enumerate}

We summarise these results in Fig.~\ref{fig:pie}. The majority of the fibrils in our sample belong to categories (b) and (c) in which the bright structures have slight differences in different lines. Considering that \CaIIIR, \CaIIK\ and \Halpha\ form close to each other in the chromosphere \citep{1981ApJS...45..635V} this can indicate that the bright fibrils are tightly packed on top of each other. In addition, the fact that the fibrils in category (b) do not fully overlap with their corresponding structures in other lines can indicate a curved morphology of the fibrils with respect to the vertical direction, which makes the different segments of the fibrils to be visible in different chromospheric lines.

Figure~\ref{fig:category} shows four examples of different \CaIIK\ fibrils and their backgrounds labelled with f1, f2, f3 and f4. The f1 and f2 cases do not appear bright in other lines which is also clear from the average intensity profiles of \CaIIIR\ and \Halpha. These fibrils have smaller $\Delta \bar{I}_{\mathrm{CaK}}$ than f3 and f4 cases. The f3 fibril (which we also studied in Fig.~\ref{fig:cak}) is bright in all three lines and overlap nearly fully with its corresponding structures in \CaIIIR\ and \Halpha. The \CaIIK\ fibril labelled with f4 also have similar bright structure in other lines as well but with a smaller intensity difference in \CaIIIR\ and \Halpha. This fibril is brighter in the wings of \CaIIK\ fibril as well. However, in other examples in Fig.~\ref{fig:category} \CaIIK\ fibrils are only brighter than their backgrounds between the K${_1}$ profile minima; this is the case for 73\% of all the fibrils in our sample.

\subsection{Inversions}
\label{sb:inversion}

\subsubsection{Methods}
\label{sbb:methods}

We applied the MPI-parallel STockholm inversion Code (STiC; \citealt{jaime16, jaime19}) to retrieve the physical parameters of the atmosphere in our sample of the bright \CaIIK\ fibrils. STiC is based on the modified version of RH radiative transfer code \citep{RH} and uses the cubic Bezier solvers to solve the polarised radiative transfer equation \citep{jaime2013}. It can fit multiple spectral lines simultaneously (in our case the \CaIIK\ line, the \FeI~6301-6302~\AA\ lines, and \CaIIIR) in statistical-equilibrium non-LTE. It takes into account partial redistribution effects (PRD) using the fast approximation \citep{jorrit12}. STiC fits the intensity in each pixel assuming a plane-parallel atmosphere (also called the 1.5D approximation). The equation-of-state utilised in STiC is obtained from the library functions in SME package code \citep{SMEcode}.

To get the best setup for the inversions, we started with running the inversions on the specific fibril and its background that we investigated in Sect.~\ref{sb:cak_profile} and is labelled with f3 in Fig.~\ref{fig:category}. This bright fibril is one of the longest and brightest ones in our sample and is a \textit{b}-category, meaning that it has a roughly similar appearance in the \Halpha\ and \CaIIIR\ data. To initialise our inversions, we used a FAL-C model \citep{VAL-C, FALIII}. We randomised the inversions to make sure that we have reached the global minimum of $\chi^2$ of the synthesised spectral profiles. We ran the inversions with 9 nodes for temperature, and 4 nodes for the line-of-sight and microtubulent velocities, distributed non-equidistantly in the range of $-7<\mathrm{\logt}<1$.

We inverted only Stokes $I$ in the \CaIIK\ line but we included all the Stokes parameters in \FeI~6301 \&~6302 \AA\ and \CaIIIR\ lines to reproduce the Zeeman broadening of the lines and keep the microtubulence as low as possible. Since we are mainly focussing on the temperature in this study, we only specified two nodes for the longitudinal component of the magnetic field, and one node for the horizontal component and the azimuth. We used the same node distribution and fitted the lines in the same way for all the inversion runs in this study.

Inverting the region marked with dotted lines in Fig.~\ref{fig:path} that is, the overlap of the fields of view of CRISP and CHROMIS (referred to as FoV hereafter), using the FAL-C model atmosphere and the above set-up is computationally time consuming because the initial atmosphere model is far from the final state and we require five randomisations in order to get a best-fit solution. Therefore, first we ran STiC on the RoI marked with the solid-lined box in Fig.~\ref{fig:path} with the same set-up as mentioned. Then, we used these inversion results to train a neural network (NN) to retrieve the mapping between the Stokes parameters of the synthesised line profiles and the physical quantities of the output atmosphere. Using the NN, we created initial model atmospheres for all the pixels in the FoV. 

We used the methods and code presented in \citet{Asensio2019NN} to create the NN.
Producing the model atmosphere for each pixel using the NN is about 1 million times faster than actually inverting it \citep{Asensio2019NN}.
For our purpose, we have used the available code with same specifications presented by \citet{Asensio2019NN}, modifying the size of the input of the NN to match our wavelength range. 
The NN outputs a model atmosphere at seven optical depths (i.e. $\mathrm{\logt = [-7,-6,-5,-4,-3,-2,-1, +0, +1]}$). The physical quantities are then interpolated to match the original depth grid of the model. Although the atmosphere model created by the NN is quite accurate at the RoI and comparable to the results of the inversions, considering that the training set is only one patch in a specific location in the FoV, we used the NN atmosphere model only as an initial guess for inverting the FoV.
%
%
With this method, we were already close to the solution so we could converge to good fits of the line profiles by applying no randomisation to the initial model, which speeded up the inversions about five times faster than using FAL-C model atmosphere with five randomisations. The inversion of the FoV (i.e. 1,604,960 pixels) took 409,746 CPU hours using Intel\textsuperscript{\sffamily\textregistered} Xenon\textsuperscript{\sffamily\textregistered} Gold 6130 processors that have a 2.10~GHz base frequency. 


\subsubsection{\CaIIK\ response functions}
\label{sbb:rf}

\begin{figure*}
  \centering
  \includegraphics[width=\linewidth]{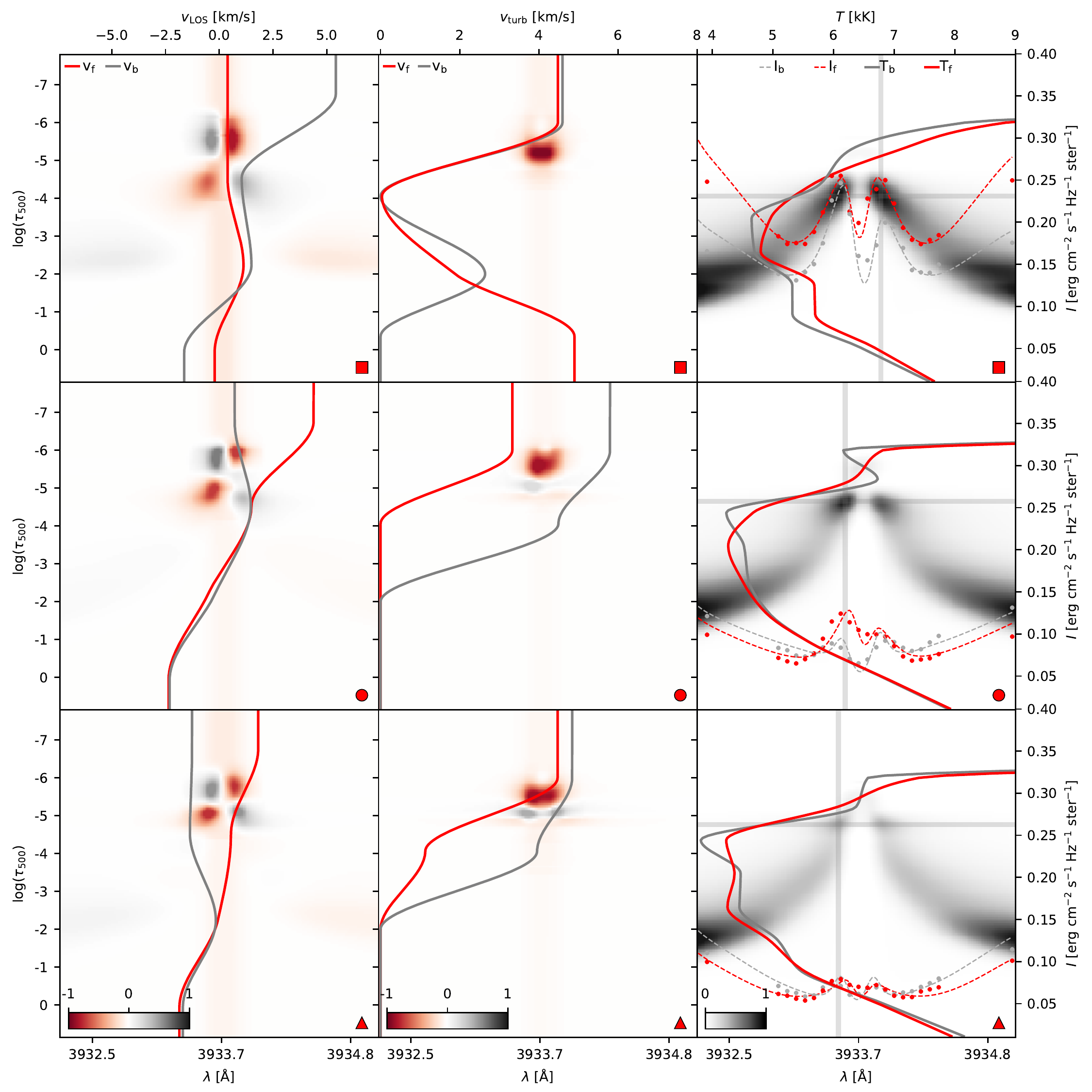}
  \caption{Normalised response function of the \CaIIK\ line to the perturbations in $v_{\mathrm{LOS}}$ (\textit{left}), $v_\mathrm{turb}$ (\textit{middle}) and T (\textit{right}) for the three points along the fibril shown in Fig.~\ref{fig:fb}; The head (\textit{top row}), midway (\textit{middle row}) and tail (\textit{bottom row}) point of the fibril are marked with a circl, square and triangle as in Figs.~\ref{fig:cak} \&\ \ref{fig:fb}. The inferred atmospheric quantities are shown with a red solid curve for the fibril, and a grey solid curve for the background lane. The observed intensities are shown with red and grey filled circles in the right column, and the fitted line profiles are shown with dashed curves. The vertical and horizontal grey narrow bands mark the wavelength point and the continuum optical depth at which the temperature response function close to the line core has a maximum.}
	\label{fig:rf}
  \end{figure*}

In order to determine the precise range of the optical depths in which the bright fibrils reveal themselves, we calculated response functions at three different spatial points along the fibril discussed in Section~\ref{sbb:fibril} following the definition of the response function for the physical parameter $X$, as $\mathrm{RF}_X(\tau, \lambda) = \delta I(\lambda)/\delta X(\tau)$ \citep{beckers75}. Since at the spatial location of this set of structures, there are no strong magnetic patches, the value of the magnetic field response function is negligible, especially around the line-core \citep{joao2018}. Therefore, we only show the response to changes in temperature, line-of-sight velocity and microtubulence. The results are presented in Fig.~\ref{fig:rf}.

\paragraph{RF$\mathrm{_{T}}$:} The \CaIIK\ line is mostly sensitive in the range of $\mathrm{\logt = [ -2,-6 ]}$. The results of the temperature response function show that the fibrillar temperature is most sensitive at the wavelength positions of the K$_{2}$ emission peaks. Moreover, the optical depth at RF$\mathrm{_{T,max}}$ is roughly the same depth where the temperature difference between this fibril and its background reaches a maximum of $\sim$~300~K (see Sect.~\ref{sbb:fibril}). This highly-sensitive depth shifts to the higher layers in the tail-point of the fibril.
In addition, the value of the response function at the K$_{2}$ emission peaks considerably decreases in the tail of the fibril.

\paragraph{RF$\mathrm{_{v_{LOS}}}$:} At the head of the fibril, the response function has the highest values within $-4<\mathrm{\logt<-6}$. This layer of highest sensitivity corresponds to the wavelength positions of the \CaIIK$_{2}$ peaks and the line core. At the midpoint of the fibril, the range in which there is the highest $\mathrm{RF_{v_{LOS}}}$ near the wavelength centre, shrinks and shifts to slightly higher layers. This behaviour is similar to the temperature response function, except that here, the value of the $\mathrm{RF_{v_{LOS}}}$ around the line-core remains almost constant in all the three points along the fibril.

\paragraph{RF$\mathrm{_{v_{turb}}}$:} The highest value of the microtubulence response function happens to be at the line-core of \CaIIK\ and in $\mathrm{\logt\sim-5}$, where also the $v_\mathrm{turb}$ peaks in both fibril and background. The region with the highest response to microturbulence shifts to $\mathrm{\logt\sim-5.5}$ at the midpoint of the fibril and its tail where the \CaIIK\ intensity difference between the fibril and its background decreases.
  
\subsubsection{Inversion of the FoV}
\label{sbb:FOV}

\begin{figure*}
  \centering
  \includegraphics[width=\linewidth]{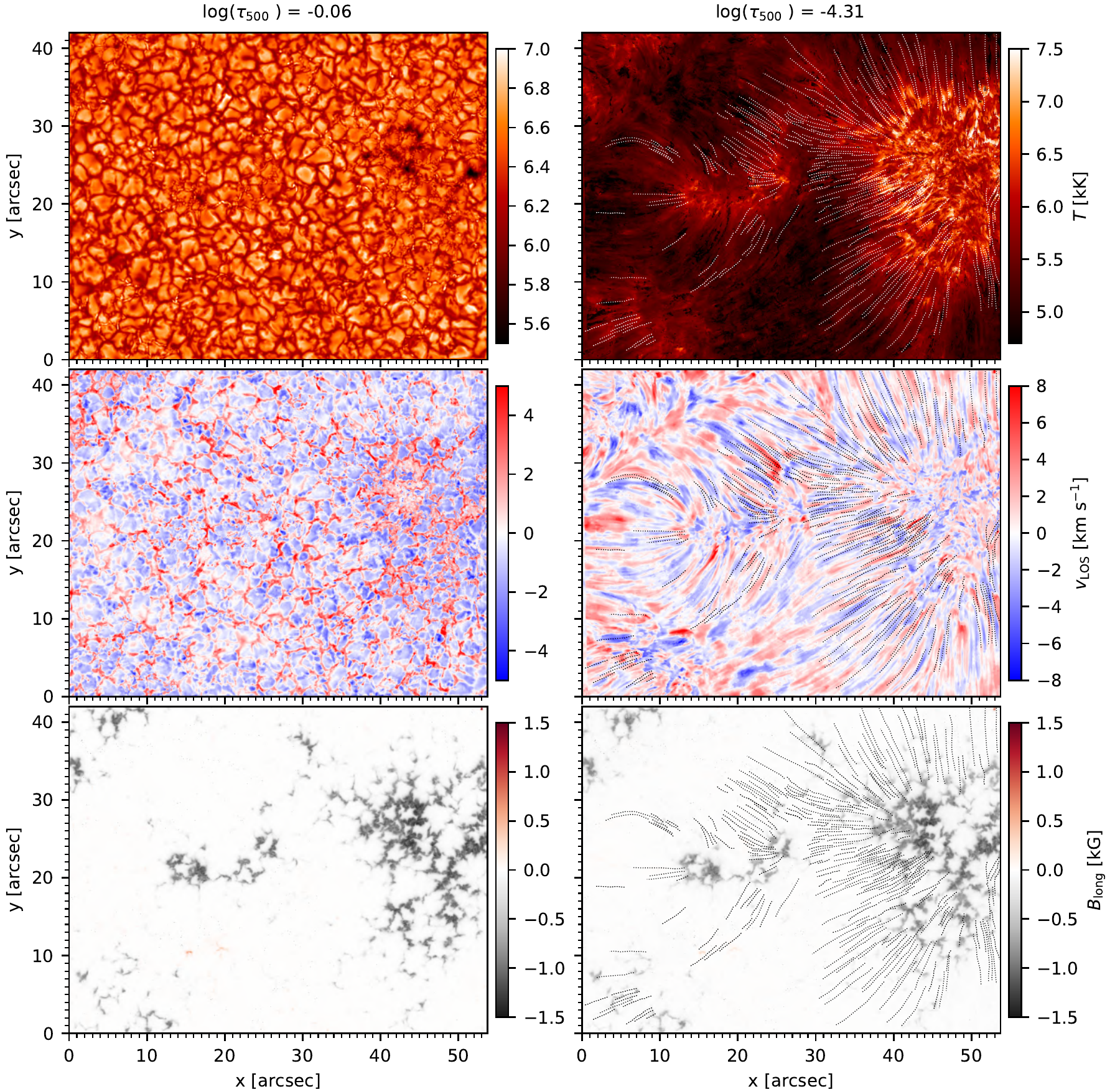}
  \caption{Inversion results of the FoV at $\logt = -0.06$ (\textit{left}) and at $\logt = -4.3$ (\textit{right}); The \textit{top} panels show the temperature and the \textit{middle} and \textit{bottom} panels demonstrate the line-of-sight velocity and the longitudinal component of the magnetic field, respectively. The paths of the \CaIIK\ bright fibrils of our sample are overplotted with dotted curves in the chromospheric panels on the \textit{right} side of the figure.}
	\label{fig:FOV_inv}
  \end{figure*}

\begin{figure*}
  \centering
  \includegraphics[width=\linewidth]{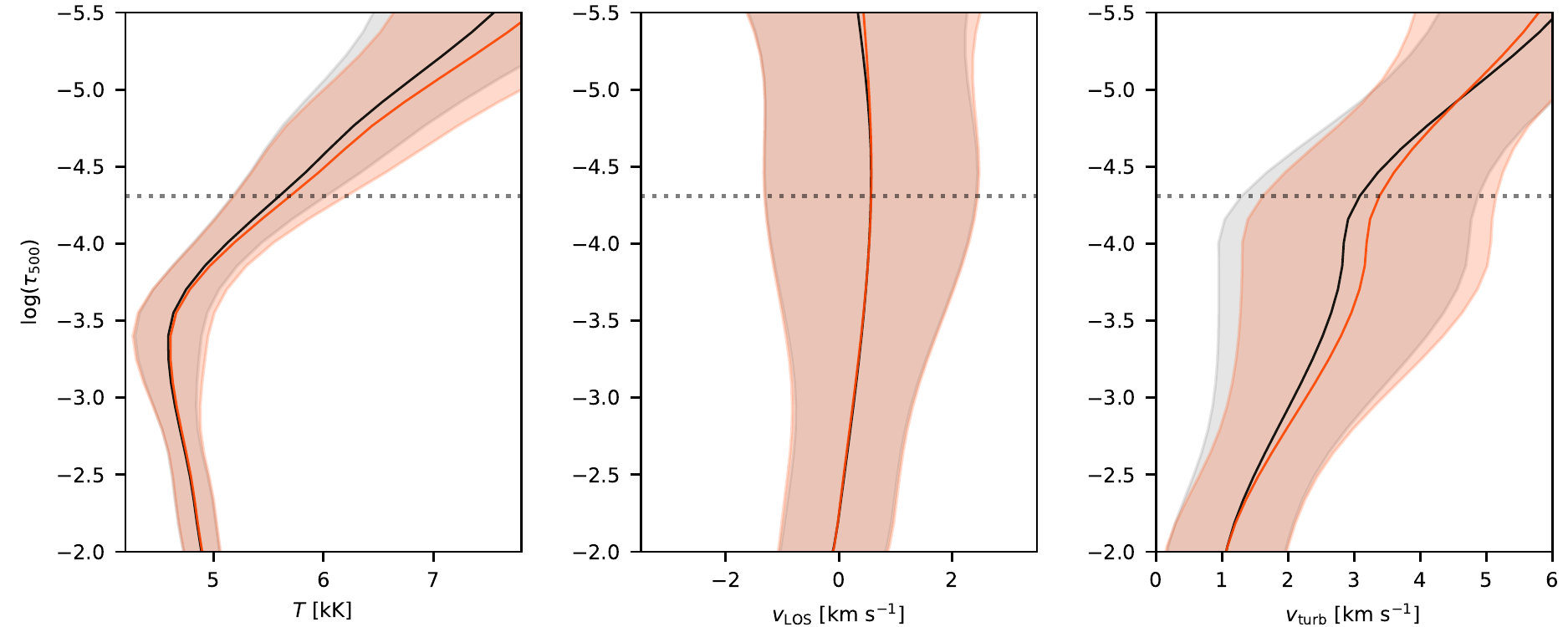}
  \includegraphics[width=0.33\linewidth]{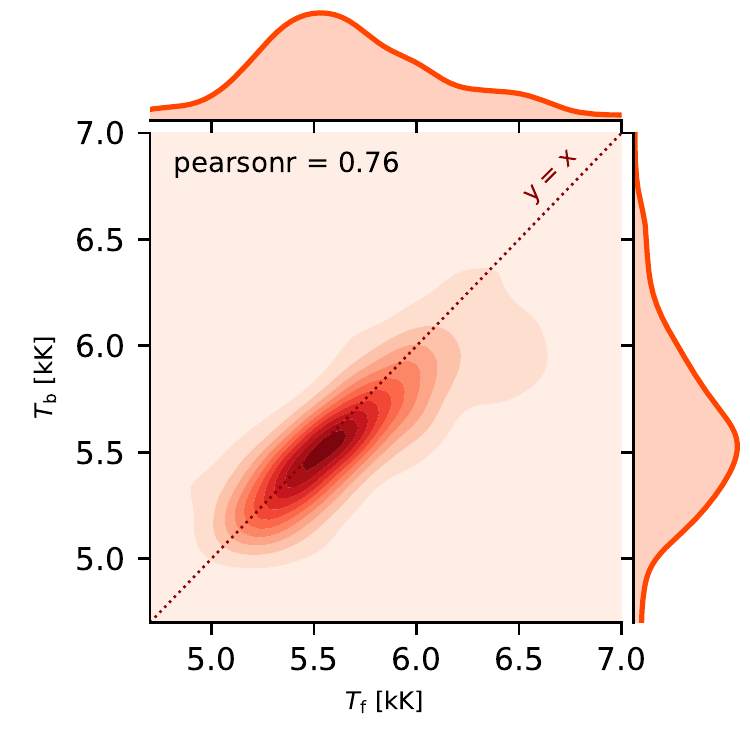}
  \includegraphics[width=0.33\linewidth]{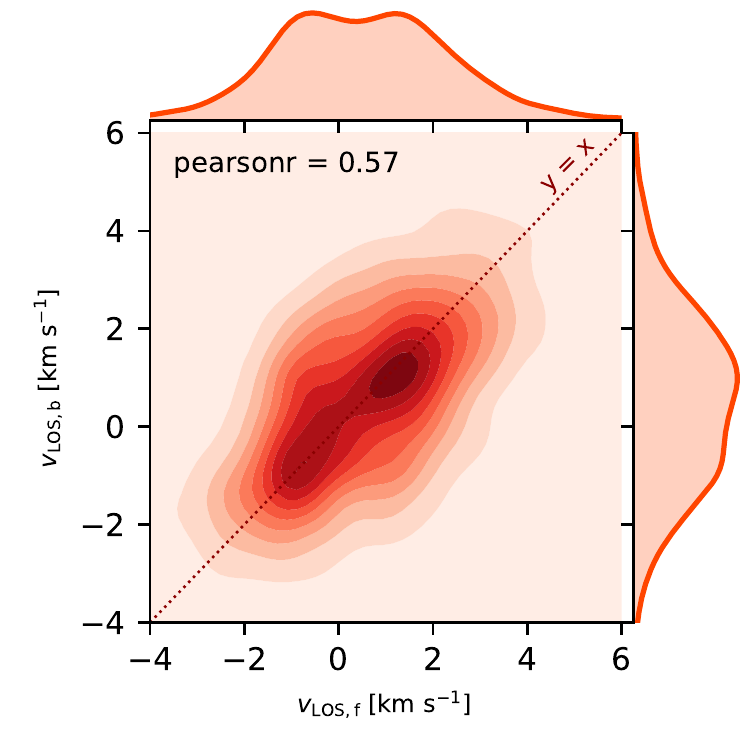}
  \includegraphics[width=0.33\linewidth]{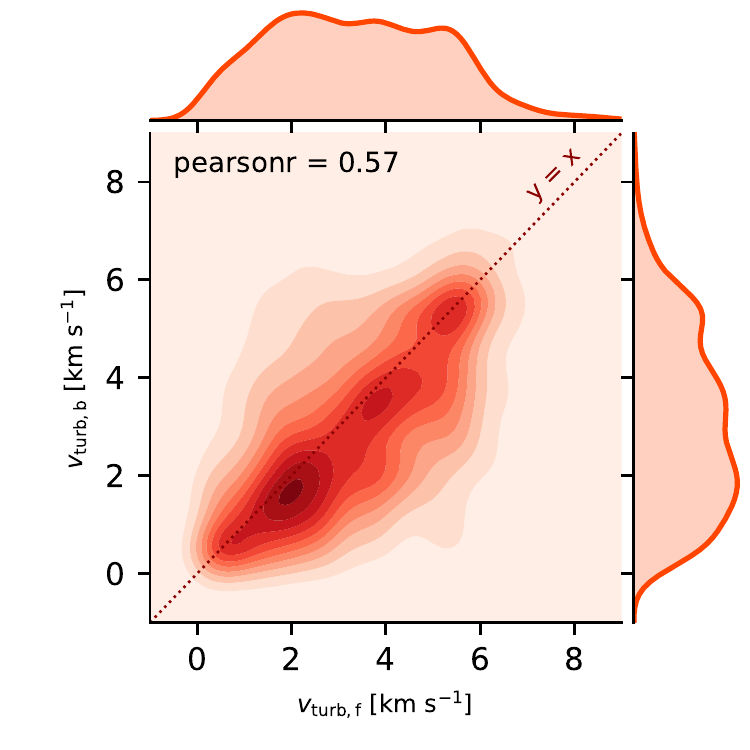}
   \caption{Inversion results of the physical quantities at the location of the fibrils in our sample and their background lanes; The \textit{top} panels show the mean profile of the temperature (\textit{left}), line-of-sight velocity (\textit{middle}) and microturbulence (\textit{right}). The fibrillar profiles are plotted with red and the background profiles are plotted with black curves. The range of $[\mu-\sigma, \mu+\sigma]$ is shown with the bands with the same colour as the mean profiles in each of the panels on \textit{top}, where $\mu$ is the mean value and $\sigma$ is the standard deviation of the physical parameters for all the pixels along the fibrils and their backgrounds. The horizontal dotted line marks the depth $\mathrm{\logt} = -4.31$. The \textit{bottom} panels, show the joint probability distributions of the physical parameters in the fibrillar pixels and their background lane at $\mathrm{\logt} = -4.31$. The Pearson correlation coefficient $r$ and the $y = x$ line is shown in each panel of the \textit{bottom} row.
       }
     \label{fig:FOV_hist}
\end{figure*}

Here, to have an overview of the physical quantities in our data especially at the location of the fibrils in our sample, we present the inversion results of the FoV with a size of $54\arcsec\times42\arcsec$, marked with dotted lines in Fig.~\ref{fig:path}. The results of the inversion of the FoV are displayed in Fig.~\ref{fig:FOV_inv} in the photosphere at $\mathrm{\logt = -0.06}$ as well as in chromosphere at $\mathrm{\logt = -4.31}$, where the fibrils have higher temperature compared to their background lanes and the \K\ peaks of the \CaIIK\ line are quite sensitive to perturbations in the physical parameters of our interest (see Fig.~\ref{fig:rf}).

\paragraph{Photosphere:}
The granulation pattern in the temperature results of the inversion in photosphere is compatible with the continuum intensity map in Fig.~\ref{fig:FOV}. The top of the granules have temperature of $\sim 6.5$~kK on average and the intergranular lanes have temperature of $\sim$~6~kK. The pores appear with a temperature of about 5.5~kK and there are temperature enhancements at the bright points in the intergranular lanes up to 7~kK. The photospheric $v_{\mathrm{LOS}}$ map resembles the granulation layout as well. The upflows at the top of the granules and downflows in the intergranular lanes range between $[-5,5]$~\kms. The pore regions have almost zero line-of-sight velocity \citep{hirzberger2003}. The longitudinal component of the magnetic field ($\mathrm{B_{long}}$) in the photosphere, agrees with the circular polarisation map in Fig.~\ref{fig:FOV} even in the case of the small-scale magnetic patches of positive polarity.

\paragraph{Chromosphere:}
The chromospheric temperature structure in Fig.~\ref{fig:FOV_inv} is in fair agreement with the \CaIIK\ and \CaIIIR\ line core intensity maps shown in Fig.~\ref{fig:FOV}. There are regions with temperatures higher than average, which are co-spatial with areas of strong photospheric magnetic field. Since the temperature at this height varies from $\sim 4$~kK to about 9~kK, the stripes of the temperature enhancement is not quite visible everywhere in the FoV. However, this thready pattern happens to trace the fibrils in many cases. It is more clear around the plage region on the right, where the temperature reaches up to 9~kK at the head segment of the fibrils (the contrast of the chromospheric temperature map in Fig~\ref{fig:FOV_inv} is clipped to include the cooler structures as well). In general the fibrils that are closer to the stronger magnetic concentrations have higher temperatures which makes them distinguishable from their nearby regions with less temperatures. The temperature enhancement in the fibrils becomes nearly washed out in the surrounding background with temperatures of $\sim$~5.5~kK as we move to the tail of fibrils. The elongated structures of temperature enhancements at the fibrillar locations are better seen in the RoI maps discussed in Sect.~\ref{sbb:ROI}.

The chromospheric line-of-sight velocity in Fig.~\ref{fig:FOV_inv} shows a roughly uniform stripy pattern of upflows and downflows which follow the overall direction of the fibrils except for the regions cospatial with magnetic concentrations; at these regions, the patches of upflow and downflow have arbitrary outlines. Here we present only the inversion results at one specific height in the chromosphere where the \CaIIK\ line is most sensitive to perturbations but the general outlook of the v$_{\mathrm{LOS}}$ remains the same along the depths at which \CaIIK$_{2}$ peaks are responsive. Even though the length and width of the stripes are comparable to those of the fibrils in many cases, there seems to be no clear correlation between the fibrillar structures as they are seen in intensity maps and the v$_{\mathrm{LOS}}$ of the chromospheric flows. However, there are few examples where the fibrils are fully located on an upflow or downflow stripe in which the absolute value of the velocity is around 4~\kms\ or higher. There are also cases in which the head segment of the fibril overlaps with a downflow patch and the rest of the fibril lies on an upflow structure. The reversed scenario is also seen along the path of a number of fibrils. There is an interesting stripe of strong downflow of $\sim8$~\kms\ originating at $(x,y) = (17\arcsec,19\arcsec)$ where there is no corresponding bright fibril detected in the \CaIIK\ wavelength-summed map; yet, there is a quite bright fibril at the exact location in the \CaIIIR\ line core. This can show that not all the structures seen in this height of the chromosphere correspond to \CaIIK\ features.

Considering the very few nodes we have specified for the magnetic properties in our inversions (e.g. only two nodes for the longitudinal component of the magnetic field), the $B_{\mathrm{long}}$ at $\logt = -4.3$, displayed in Fig.~\ref{fig:FOV_inv}, is not showing much difference from its map at the photosphere.
The negative polarity of the $B_\mathrm{long}$, however, reaches only up to $-1$~kG. In addition, the small-scale patches of positive polarity, existing in the photosphere, are almost vanished in the chromosphere. In summary, all the fibrillar structures are emanating from the magnetic concentrations.

\paragraph{Statistics of the physical parameters:}

Based on the results of the inversions of the FoV, we extracted the physical parameters of all the pixels along the bright fibrils in our sample and their background. The results are shown in Figure~\ref{fig:FOV_hist}. The average profile of temperature for the depth range in which the \CaIIK\ line is sensitive to perturbations (see Fig.~\ref{fig:rf}), show a very similar behaviour in both fibrils and their backgrounds; they overlap below $\logt=-3.5$. Then the fibrillar average temperature profile decouples and rises with a larger gradient. This temperature difference between their average profile keeps increasing up to $\sim400$~K at $\logt = -5.5$. The joint probability distribution of the fibrillar temperature and the background at the height where the response function peaks at the \K\ peaks wavelength positions (i.e. $\logt = -4.31$) demonstrates the similarity between the fibrils and their backgrounds. This small temperature difference was also expected from panel b in Fig.~\ref{fig:k2}.

The average profile of the line-of-sight velocity in the fibrils and their background overlaps completely along all the heights in the range of $\logt = [-2,-5.5]$. The velocity distribution resembles the study of the \K-peak asymmetry in Fig.~\ref{fig:k2}c, showing similar velocity structures in both fibrils and backgrounds with no preference for upflows and downflows. 

The average profile of the microturbulence in the fibrils is slightly higher within the range of $\logt = [-2.5,-4.8]$ with a maximum difference of $\sim 0.5$~\kms\ around $\logt = -3.8$. The corresponding distribution shows that there is also a local correlation between the microturbulence in the fibrils and their background: the microturbulence in the local background tends to be somewhat lower than in the fibrils. This is consistent with the results of the \K~separation study in Sect.~\ref{sb:cak_profile}.

\subsubsection{Region of interest}
\label{sbb:ROI}

\begin{figure*}
  \centering
  \includegraphics[width=\linewidth]{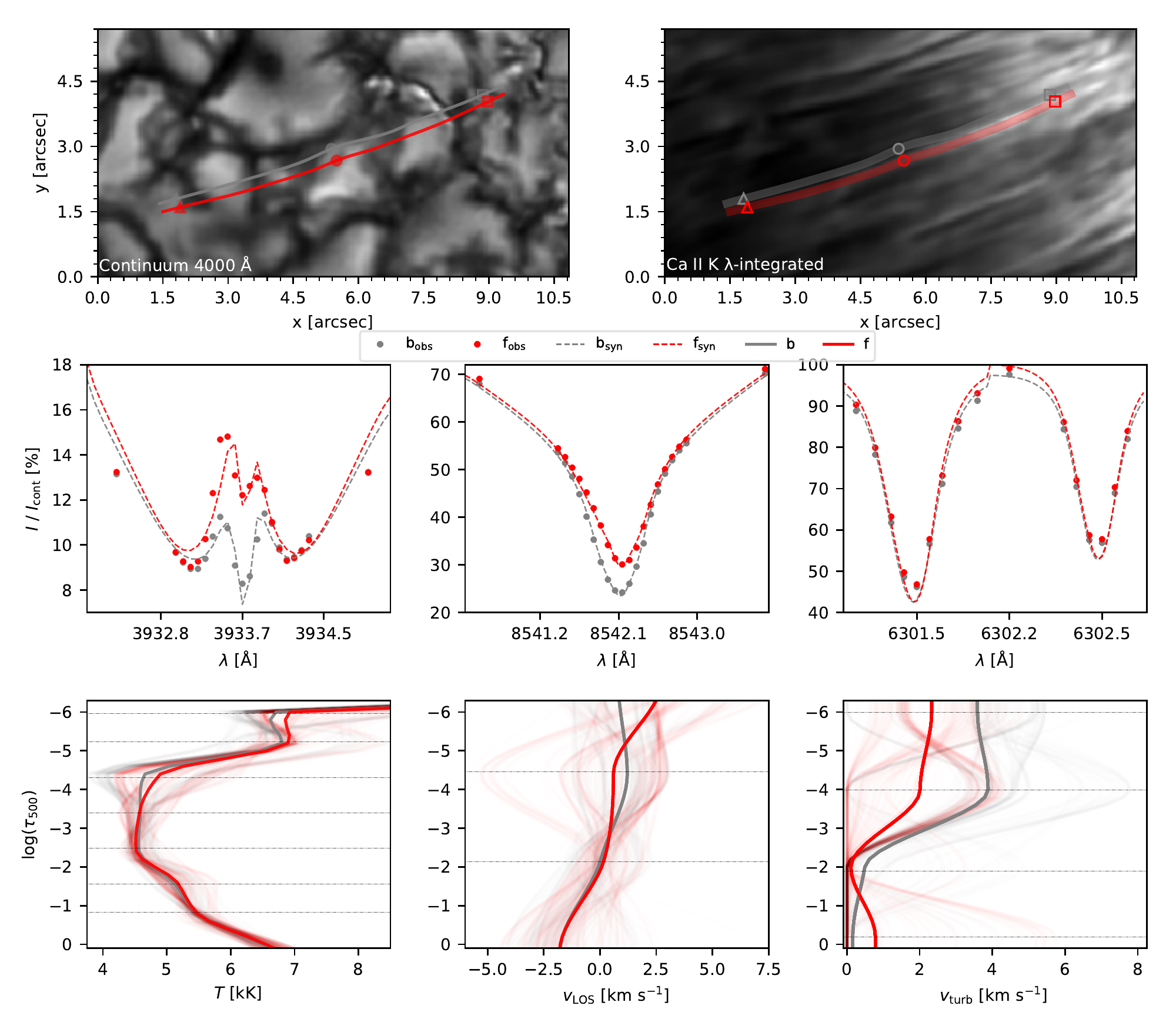}
  \caption{Inversion results along the fibril--background set labelled with f3 in Fig.~\ref{fig:category}; \textit{Top} panels show the fibril (red) and the background (grey) overplotted on intensity maps of the RoI marked with a solid-lined box in Fig.~\ref{fig:FOV} in the continuum and \CaIIK\ summed light. The head, midpoint and tail of the structures are marked similarly to Fig.~\ref{fig:cak}. The average observed and synthesised line profiles along the structures are shown in the \textit{middle row} panels. The inversion results of the temperature, line-of-sight and microturbulence velocities for each individual pixel along the path of the structures are plotted with the washed-out lines and the average profiles are shown with the bold lines in the \textit{bottom} panels. An animated version of this figure showing the fit of each indiviudual line profile along the fibril is available online.}
	\label{fig:fb}
  \end{figure*}

\begin{figure*}
 \begin{adjustbox}{addcode={\begin{minipage}{\width}}{\caption{
   Temperature (\textit{left}), line-of-sight velocity (\textit{middle}) and microturbulent velocity (\textit{right}) as derived from the inversions. Each row shows a different heights in the atmosphere, as marked in the temperature panels. The identified \CaIIK\ bright fibrils are overplotted in white and the fibril studied in Sect.~\ref{sbb:fibril} is shown with a red path. The contours levels overplotted on the temperature panel at $\logt=-0.06$ show the strength of the longitudinal component of the photospheric magnetic field at 200~G (cyan) and 800~G (green).
       \label{fig:ROI}
   }\end{minipage}},rotate=270,right}
   \includegraphics[scale=.85]{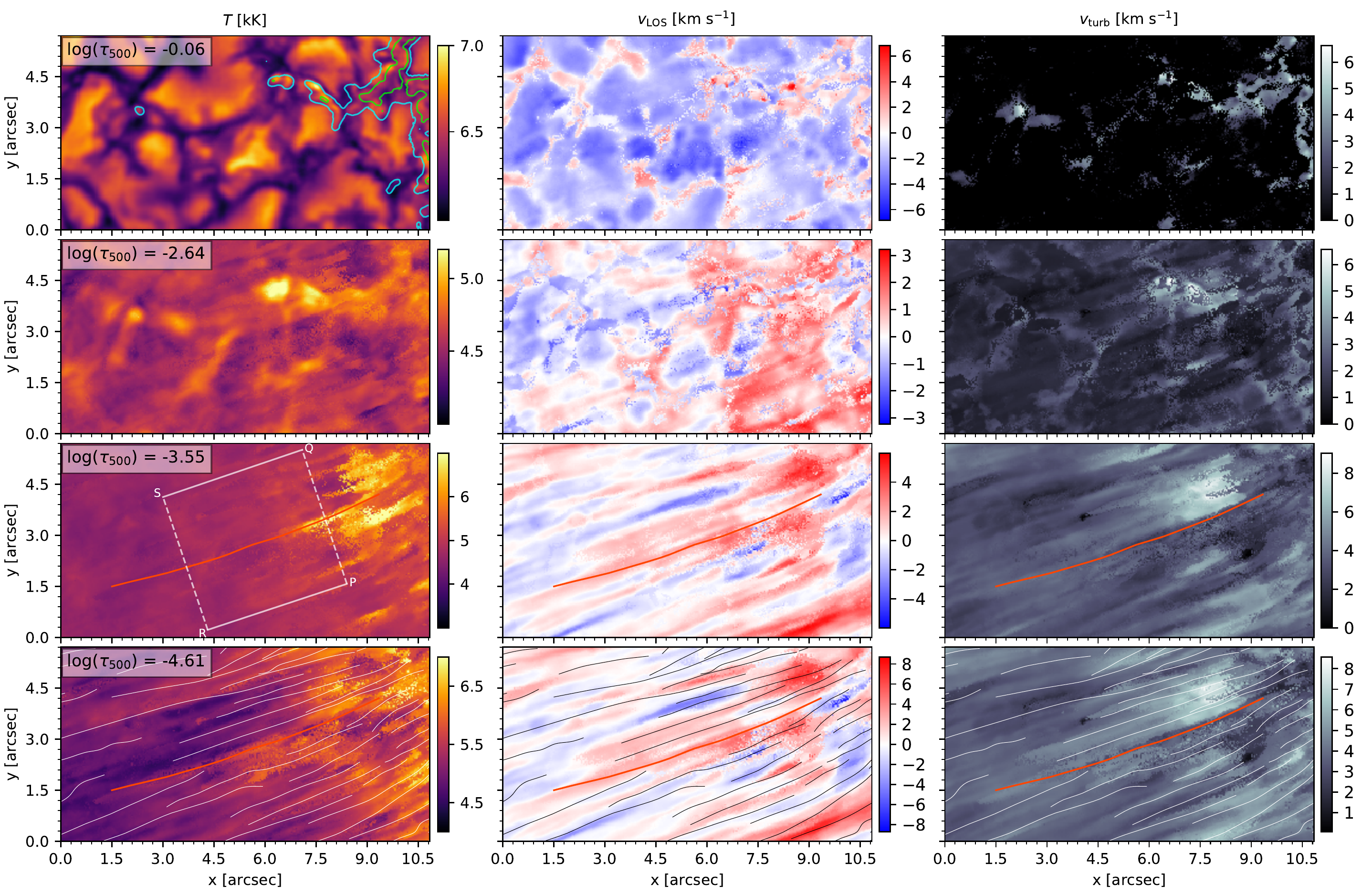}%
 \end{adjustbox}
\end{figure*}

\begin{figure*}
  \centering
  \includegraphics[width=\linewidth]{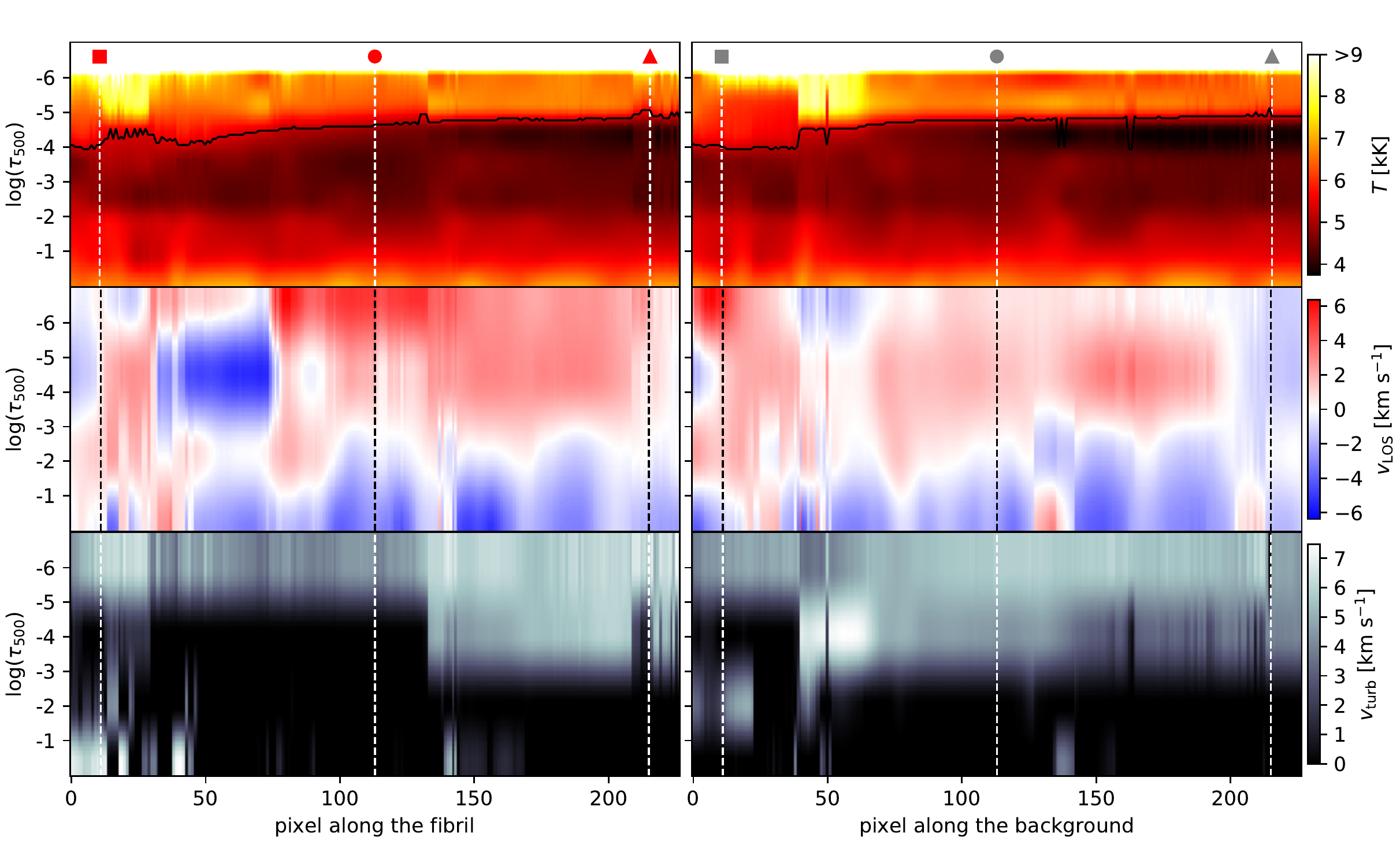}
   \caption{Atmospheric properties along the fibril (\textit{left}) shown in Fig.~\ref{fig:fb} and its background (\textit{right}). \textit{Top}: temperature; \textit{middle}: line-of-sight velocity; \textit{bottom}: microtubulent velocity. The square, circle and triangle symbols indicate the head, midway point and the tail of the fibril, respectively. The dashed lines help to guide the eye. The black curve marks the height where the response function of the \CaIIK\ line with respect to temperature has its maximum (see Fig.~\ref{fig:rf}), which is typically at the wavelengths of the \K\ peaks.}
     \label{fig:along}
\end{figure*}

\begin{figure*}
  \centering
  \includegraphics[width=\linewidth]{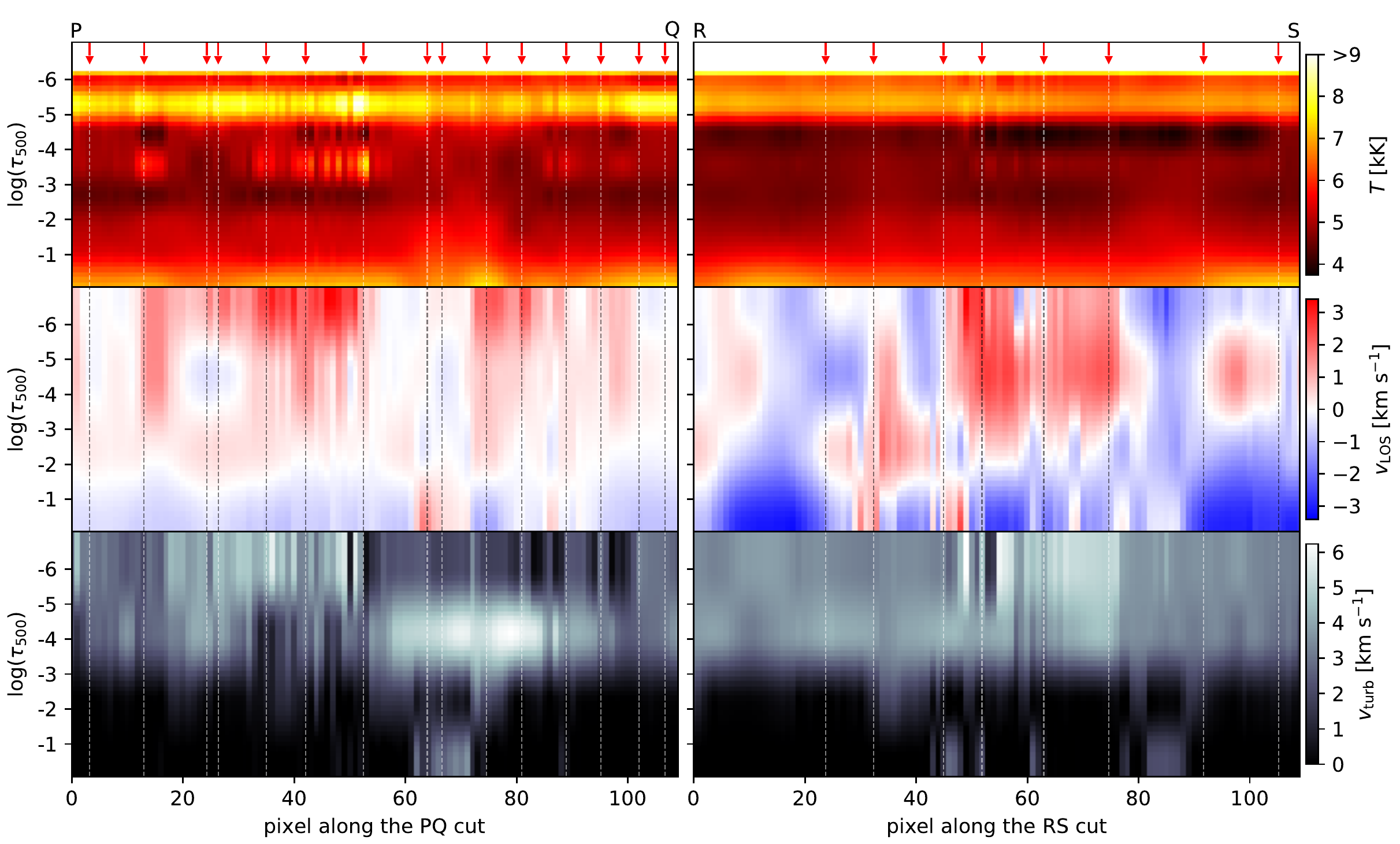}
   \caption{Inversion results of temperature (\textit{top}), line-of-sight velocity (\textit{middle}) and microtubulent velocity (\textit{bottom}) in PQ (\textit{left}) and RS (\textit{right}) cuts in the atmosphere across the head and the tail of the fibrils, respectively. The cross point of the fibrils with the cuts are shown with the red arrows. The location of the cuts and the bright fibrils pass through them are in the RoI are marked in Figures~\ref{fig:category} and \ref{fig:ROI}.
        \label{fig:across}
       }
\end{figure*}

To focus on the fine details of the physical parameters of the fibrils and their background lanes, we present the inversion results along the specific set of fibril--background which is marked with f3 in Fig.~\ref{fig:category}, and with the RoI of $10\farcs8 \times 4\farcs4$ marked with the solid-lined box in Fig.~\ref{fig:path}. The RoI contains the fibril shown in Fig.~\ref{fig:category}. We present the RoI inversion results with various approaches.

\paragraph{Individual fibril and background:}
\label{sbb:fibril}

The inversion results are shown in Fig.~\ref{fig:fb}. The fibrillar structure stretches over several granules with the head located above the disturbed granulation, where some magnetic concentrations are located. On average, the inversion fitted the line profiles well. In the case of \CaIIK, it has preserved the line width of both fibril and the background lane profiles and reproduced the reversal K$_3$ peak in the line centre.

The average temperature profiles of the fibril and its background overlap between the photosphere and lower chromosphere. At about $\logt = -3.4$ the fibril temperature decouples and starts rising with a steeper gradient than the background. This thin layer of higher average fibrillar temperature within $\mathrm{-3.4<\logt<-4.5}$ causes the \K\ peaks of the \CaIIK\ intensity profile in the fibril to be higher than the those of in the background. The maximum temperature difference between the average temperature of the fibril and its background, which is $\sim$~200~K, happens at $\mathrm{\logt=-4.3}$, where the $\mathrm{-3.4<\logt<-4.5}$ fibril temperature is 4.9~kK. The brightest part of the fibril is its head (around the square symbol). There, the temperature around $-5<\logt<-4$ (see Sect.~\ref{sbb:rf}) can be as high as $6$~kK. 

The middle panel of the bottom row shows that the microtubulent velocity along the fibril is generally lower than in the background, which is contrary to the average behaviour as shown in Fig.~\ref{fig:FOV_hist}. This behaviour is consistent with the larger \K\ peak separation in the upper half of background lane compared to the bright fibril in Fig.~\ref{fig:cak}.

The animation of the pixel-to-pixel variations of Fig~\ref{fig:fb} shows that line-of-sight velocity along the fibril and the background lane varies drastically from pixel to pixel. The \CaIIK\ intensity profiles in the brighter segment of the fibril (between the points marked with the square and the circle) have higher K$_{\mathrm{2V}}$ peaks. This can indicate upflows at about 1~Mm\,---\,1.5~Mm above the photosphere and downflows above that height \citep{carlsson97,johan18}. The overall scenario agrees with the $v_{\mathrm{LOS}}$ in most of the close-to-head pixels of the fibril with negative velocities (meaning upflow) below $\mathrm{\logt=-2.1}$ and positive velocities ranging from $-0.5$ to $-4.5$~\kms\ in the higher layers (e.g. $\mathrm{\logt = -4.5}$) at which the fibril has also higher temperature on average than its background lane. As we move towards the tail (the fainter segment of the fibril) the $v_{\mathrm{LOS}}$ tends to be negative up to $-2.5$~\kms deeper in the atmosphere.

\paragraph{Temperature in the RoI:}
The atmospheric properties for the RoI are presented for four different depths from the photosphere to the chromosphere in Fig.~\ref{fig:ROI}. At $\mathrm{\logt} = -0.06$. there is a hot spot on the edge of the granule with a temperature of $\sim$~7.5~kK at the point $\mathrm{(x,y) = (7.5\arcsec, 4.2\arcsec)}$. This spot coincides with a longitudinal magnetic feature with a strength of about 200~kG in the photosphere. The feature is located in the path of the specific fibril showed in Fig~\ref{fig:fb}. 

Higher up in the atmosphere, at $\mathrm{\logt}=-2.6$, we start to see the imprint of elongated structures in temperature. Temperature variations at this height mainly change the intensity in the wings of the \CaIIK\ line (see Fig.~\ref{fig:rf}). The head of a fibril is clearly distinguishable at $\mathrm{(x,y) = (7.9\arcsec, 4.6\arcsec)}$, with a temperature enhancement of $\sim$~300~K with respect to its surroundings. 
At this height the imprint of the enhanced intergranular lanes due to the reverse granulation is visible. In addition, there are two hot blobs with $T>5$~kK, cospatial with magnetic concentrations in the photosphere.

At $\mathrm{\logt = -3.5}$, there is no trace of the intergranular lanes nor the hot blobs in temperature. Instead, the temperature in the RoI shows elongated features that have a small temperature enhancement compared to their surroundings. The elongated high-temperature features in the top right corner of the RoI coincide with the magnetic concentrations. They appear to be the hot head parts of several fibrils. At this height the mid to end parts of the fibrils are hardly visible.

At the height $\mathrm{\logt = -4.6}$, the fibril shows a temperature of about 6.3~kK which is even higher than its head at this depth. 
At this height, also the tails of the fibrils have enhanced temperatures, contrary to the situation at $\mathrm{\logt = -3.55}$. We have seen this behaviour in the response function of the \CaIIK\ line in Sect.~\ref{sbb:rf} where we showed that K$_{2}$ peaks, at the head of the fibril that is closer to the magnetic patches, are most sensitive to the temperature perturbations in the lower depths compared to the case in the tail-point of the fibril. There are few fibrils in the bottom right corner of the $\mathrm{\logt = -3.55}$ panel with temperatures reaching to 7~kK mostly in their head segment. As it can be seen from the panels of $\mathrm{\logt = -2.64, -3.55, -4.61}$, not all the fibrils appear in all the depths in which \CaIIK\ is sensitive to temperature perturbations. The imprint of the fibrils, appearing as stripes of slight temperature enhancement, persists up to $\mathrm{\logt = -5.5}$ where they reach temperatures 
above $7$~kK even at the location of the tail of some fibrils. For the inversion results of temperature, $v_{\mathrm{LOS}}$ and $v_\mathrm{turb}$ along all the optical depths in the range of $[-0.61,0.06]$, please see the video in the online material of the article.

\paragraph{Line-of-sight Velocity in RoI:} 
The granulation pattern is clearly visible at $\mathrm{\logt}=-0.06$ with upflows in the granules and downflows in the intergranular lanes. This pattern gets disturbed as the stripes come into sight at $\mathrm{\logt}=-2.64$. Higher up at $\mathrm{\logt}=-3.55$ the v$_{\mathrm{LOS}}$ structure turns to a zebra pattern of upflows and downflows. The overall pattern is preserved at $\mathrm{\logt}=-4.61$ as well but the absolute value of the line-of-sight velocity increases. This value increases up to about 10~\kms\ with the same pattern as we move higher in the atmosphere until we reach $\mathrm{\logt = -6}$ (not shown in Fig.~\ref{fig:ROI}). We note that although the specific fibril (overplotted in red) is elongated where there is a downflow stripe of $\sim2$~\kms\ along its path, looking at the 2D maps of $v_{\mathrm{LOS}}$ at different heights suggests that generally the fibrils have no preference to happen at the location of upflows nor downflows.

\paragraph{Microturbulent velocity in RoI:} The results of the microturbulence in Fig.~\ref{fig:ROI} show that almost everywhere over the RoI at $\mathrm{\logt=-0.06}$, $v_\mathrm{turb}$ is close to zero, except for the regions where there are magnetic concentrations and in some intergranular lanes. At these locations the value of $v_\mathrm{turb}$ can be up to 6~\kms. At $\mathrm{\logt=-2.64}$, the range of the $v_\mathrm{turb}$ variations remains the same. The intergranular lanes have a higher $v_\mathrm{turb}$ than the granules. On top of this pattern, there is a weak, but clearly visible, imprint of the fibrillar structure.

The average microturbulence increases higher up in the atmosphere with a similar pattern at both $\mathrm{\logt}=-3.55$ and $-4.61$. Its value stays below 6~\kms\ in most of the regions in the RoI except for a region in around $\mathrm{(x,y) = (7.8\arcsec, 4\arcsec)}$. Some fibrils have a higher microturbulent velocity than the background, and some do not. The statistics demonstrated in Fig~\ref{fig:FOV_hist}, show that fibrils tend to have a slightly higher $v_\mathrm{turb}$ on average.

We note that since we used the wavelength-summed \CaIIK\ data to identify the fibrils, we are not including most faint structures that only appear bright in one wing of the \CaIIK\ profile since they get washed out in the summed maps.

\paragraph{Atmosphere cut -- \textit{Along the fibril and background:}}

To further investigate the behaviour of the fibrils compared to their nearby dark background lane, we extracted the inversion results along the paths of the fibril shown in Fig.~\ref{fig:fb}. The latter shown in Fig.~\ref{fig:along}. 
The temperature drops to its minimum point with a value of $\sim$~5~kK in the locations close to the head of the structures (marked with squares) and $\sim$~4~kK at the locations closer to the tail (marked with triangles). The depth at which \CaIIK\ is most sensitive to temperature (RF$\mathrm{_{cak,max}}$; shown by the black line) is located just above the onset of the chromospheric temperature rise. We note that the temperature difference between the fibril and the background lane, at the height of RF$\mathrm{_{cak,max}}$, is of the order of 200--300~K, which is hard to see in this figure. The fact that the \CaIIK\ response function peaks at larger optical depths close to the head of the fibril (also see Fig.~\ref{fig:rf}) could be an indicator of the curved morphology of the fibrils on a geometrical height scale. However, since the inversion results are on an optical depth scale only, we cannot prove this.

The line-of-sight velocity variations along the cuts display the granules as blobs of upflows where $\mathrm{\logt > -2}$. Above this height, both the fibril and the background lane show downflows of about $\sim 3$~\kms, except for an upflow of $\sim 6$~\kms\ in the fibril between pixel 30 and 70. This blob is cospatial with the brightest segment of the fibril in \CaIIK\ summed map (see Fig.~\ref{fig:fb}, top panels). At the corresponding location in the background lane, there is only a weak downflow.

The microturbulence along both the fibril and its background in the bottom row of Fig.~\ref{fig:along}, show consistently low microturbulence (less than 2~\kms) where $\mathrm{\logt > -3}$. In the first 130 pixels of the fibril length, v$_{\mathrm{turb}}$ remains low through to the height of $\mathrm{\logt =-5}$. Between the layer in which the {\CaIIK$_{2}$} peaks are sensitive to perturbations in microturbulence ($\mathrm{-5<\logt<-6}$), it starts rising to more than 4~\kms. However, in the last 95 pixels of the length of the fibril, corresponding to its tail, $v_\mathrm{turb}$ starts rising from a lower depth of $\mathrm{\logt\approx-3}$. We note that the K$_{2}$ wavelength separation in the tail of this fibril is larger than the wavelenght separation close to the head (see Fig.~\ref{fig:cak}). This is consistent with the behaviour of the microturbulence in this specific case.

\begin{figure*}
  \centering
  \includegraphics[width=.462\linewidth]{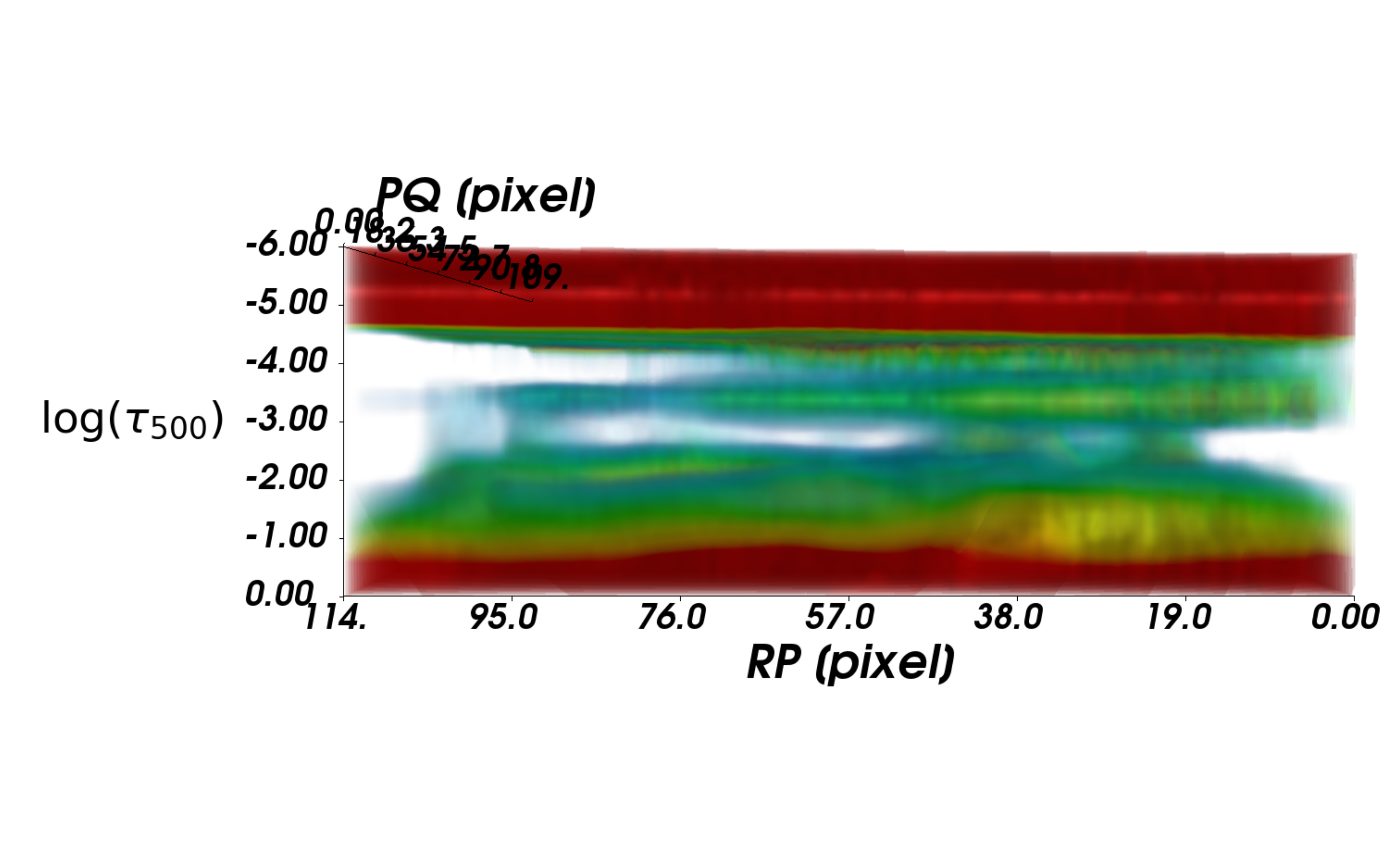}
  \includegraphics[width=.533\linewidth]{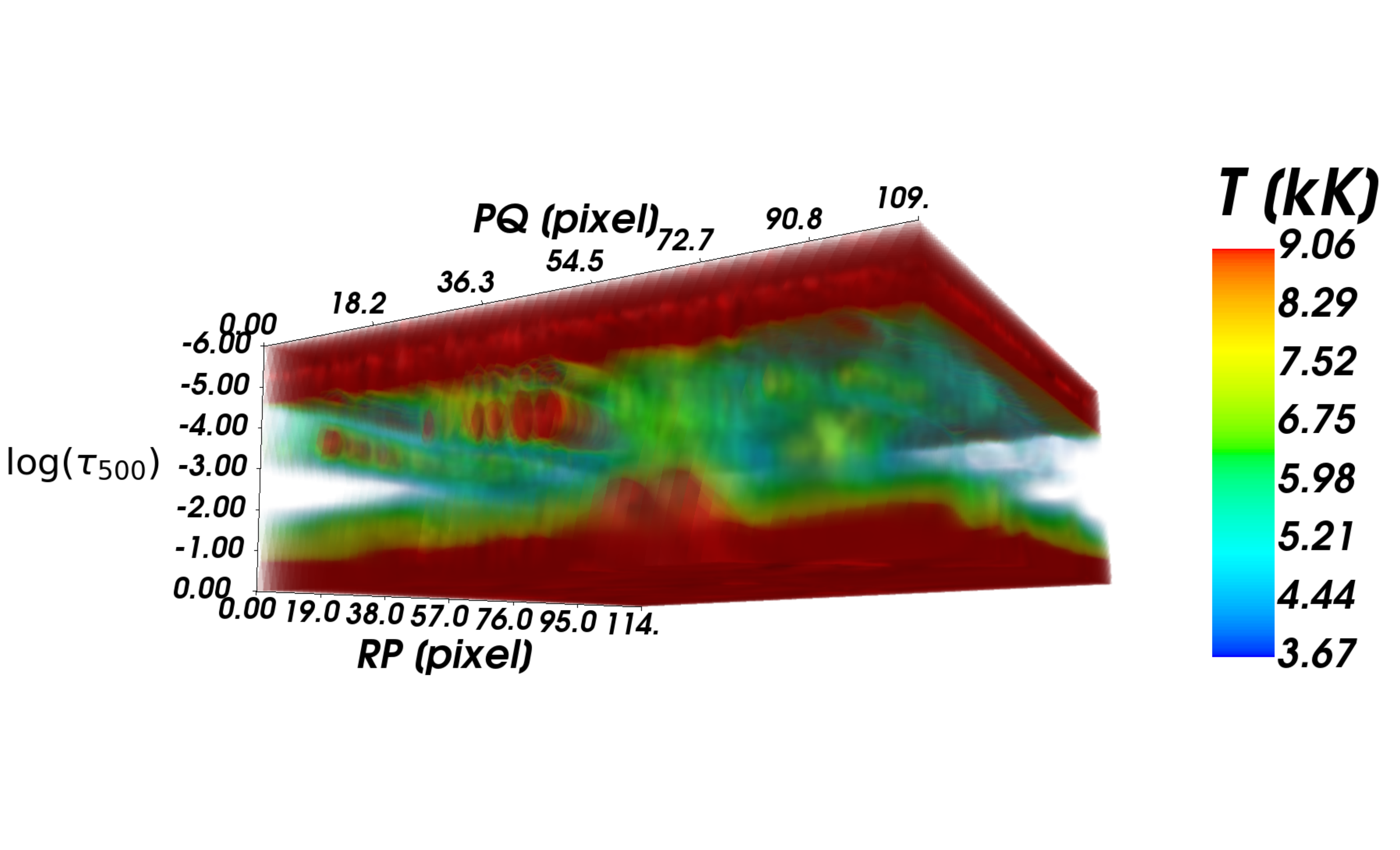}
  \caption{The 3-D visualisation of the temperature variations inside the patch PQSR along the atmosphere depth ($\mathrm{logt}$) shown in the RoI panels of Fig.~\ref{fig:category}, viewed from RP-side (\textit{left}) and from a low angle (\textit{right}). Temperatures below 3.67~kK are being transparent to make it possible to look through the cube.}
	\label{fig:cube}
  \end{figure*}

\paragraph{Atmosphere cut -- \textit{Across the structures:}}

As it was shown in Fig.~\ref{fig:fb}, the temperature difference between the fibril and its background is about $200$~K. To further illustrate this temperature enhancement, Fig.~\ref{fig:across} shows the atmospheric properties along the lines PQ and RS in Fig.~\ref{fig:category}. These lines are roughly perpendicular to the overall orientation of the fibrils in the RoI so they cut across several fibrils.

The PQ cut corresponds to the head of most of the fibrils passing through it which leads to blobs with a temperature of $6-8$~kK between $\mathrm{\logt=-3}$ and $-4$, while the rest of the atmosphere in this height layer has a temperature of 6~kK. 

The hottest blob, located at pixel 53 along the cut, belongs to the specific fibril studied in Sect.~\ref{sbb:fibril}. There are quite a few imprints of bright fibrils between pixels 12 to 53 along the PQ cut which either appear dim or are hardly visible in the \CaIIK\ observation. There are also temperature enhancements at heights between $\mathrm{\logt = -5}$ to $-5.7$ which do not exactly coincide with the fibril cross-points but appear slightly shifted. For instance, in case of the f3 fibril at pixel 53, there is a temperature enhancement between $\mathrm{\logt = -5}$ to $-5.7$ that is located at pixel 50. 
We speculate that such simultaneous nearly co-spatial temperature enhancements at those two heights in our inversion results can be connected or even be considered to be the same fibril, given that fibrils have an average width of about 200~km (\citealt{gafeira17}, which corresponds to $\sim6$~pixels in our data). However, the \CaIIK\ line sensitivity to temperature perturbations at optical depths smaller than $\logt = -5$ is low and we cannot conclude more based on our data.

The imprint of the fibrils within pixels 55 to 85 of the PQ cut seems to be blurred by the temperature rise owing to the presence of a magnetic concentration in the lower layers around $\mathrm{\logt=-2.5}$ (see Fig~\ref{fig:ROI}). The temperature enhancement in the fibrils at $\mathrm{\logt = -3}$ in the RS cut are rather weak, because this cut intersects the tail of the structures. The tail segments of the fibrils are generally less bright in \CaIIK\ observations. The temperature enhancement at these points hardly reaches 6~kK.

The line-of-sight velocity in the PQ and RS cuts in Fig.~\ref{fig:across} indicates downflows at the cross points of a number of bright structures. Nevertheless, neither upflows nor downflows seem to dominate at the points where the fibrils pass through, which is consistent with the result in Fig.~\ref{fig:FOV_hist}.

The microturbulence in the PQ cut is $> 6$~\kms around a height of $\mathrm{\logt = -4}$ between pixels 55 to 85 where the magnetic element is located. The rest of the PQ cut and the RS cut show similar behaviour: the value of $v_\mathrm{turb}$ is $\sim3$~\kms around $\mathrm{\logt = -4}$. Below this height it is typically $<1.5$~\kms, at larger heights it varies more strongly with $ 0 <v_\mathrm{turb} <6$~\kms.

\paragraph{3D visualisation of the temperature:}

To highlight the temperature difference in the fibrils compared to their surrounding background atmosphere we applied 3D visualisation of the temperature for optical depths between $\mathrm{logt}=0$ and $\mathrm{logt}=-6$ in the patch of the solar surface marked with the PQSR box in Fig.~\ref{fig:ROI}. We applied the \verb|volume| filter from the \verb|mayavi| package for Python to this cube and looked at it from different viewing angles. The result is shown for two different viewing angles in Figure~\ref{fig:cube}. 

The points with temperatures below 3.76~kK are set to full transparency and the points with temperature up to 5.45~kK are transparent with a degree proportional to their temperature value. This approach makes it possible to look through the box and highlight where there are temperature enhancements. All the visible points inside the cube, which have temperatures varying from 3.76 to 9.06~kK, are colour-coded as shown in the colourbar of the figure. The left panel of Fig.~\ref{fig:cube} shows the RP side view of the cube. The fibrils are visible in a layer around $\mathrm{\logt = -3.5}$ with a maximum temperature of $\sim$~7.5~kK close to the point P which is near the head of the existing fibrils in the box. The temperature enhancement almost vanishes close to point R which is close to the tail of the structures. This thin fibrillar layer from $\mathrm{\logt=-3}$ to $-4$ is sandwiched between two layers with lower temperature. 

The right panel of Fig.~\ref{fig:cube} shows the cube as being rotated 120$^{\circ}$ clockwise around the $\mathrm{\logt}$ axis from its position in the left panel and tilted 5$^{\circ}$ in the inward direction with respect to the paper. Looking through the cube from this angle reveals the bright fibrils as hot threads with a temperature up to 9~kK in their core and a halo with lower temperature of $\sim$~6~kK surrounding the core of this hot thread. The hot blobs of fibril cross-points at pixels 13 and 53 of the PQ cut in Fig.~\ref{fig:across} are clearly distinguishable as hot threads in the right panel of Fig.~\ref{fig:cube} as well. It is also noticeable that the structures, especially between pixels 36 to 54 in PQ side are so tightly packed that it is challenging to distinguish the bright fibrils from their cooler background atmosphere. This can explain the small value of the temperature difference between the bright structures and their background lanes. To see the different angles of the temperature cube in Fig.~\ref{fig:cube}, please see the video in the online version of the article.

\section{Conclusions and discussion}
\label{sec:con}

We presented observations with high spatial and spectral resolution including spectropolarimetric data in a plage region. The FoV in the \CaIIK\ images is covered with bright fibrils that have strong \K-peaks. 
They appear brighter than the background owing to a local temperature enhancement in the chromosphere. The typical temperature difference between the fibrils and their surroundings is about 100--200~K at $-3.5 < \logt< -4.3$ that is, where the \CaIIK$_2$-peaks are sensitive to the temperature perturbations. Because of the relatively short wavelength of the \CaIIK\ line, temperature differences of only a few percent can give rise to intensity differences well above 10\%.

The temperature enhancement in the fibrillar regions as determined using Atacama Large Millimeter Array \citep[ALMA;][]{ALMA} and \Halpha\ data is higher than what we have found \citep{molnar2019}. 
Bright fibrillar imprints have also been seen in the 3~mm-band by \citep{joao2019}, who performed simulatanous non-LTE inversion of ALMA data and \MgIIhk\ data from the Interface Region Imaging Spectrograph \citep[][IRIS]{IRIS}. It will be interesting to investigate the bright fibrils and their temperature in the higher layers of the atmosphere using ALMA and IRIS data and compare it to our findings.

We also compared the bright fibrils in \CaIIK\ to \CaIIIR\ and \Halpha\ intensity images. We found that the brighter the fibrils are in \CaIIK\ observations, the more likely they also appear bright with the same appearance in the other chromospheric lines. However, the fibrillar structures do not necessarily appear the same in these three lines, despite the fact that they form at similar heights in the chromosphere \citep{jorrit12, johan18}. This suggests that a significant fraction of fibrils have small vertical extent, and are packed not only horizontally over the solar surface but also vertically. This is also seen in the heights where the fibrillar temperature enhancements are located in our inversions: the imprint of different fibrils appears at various chromospheric heights.

Despite their clear temperature structure, the \CaIIK\ fibrils do not show any particular line-of-sight velocity structures. However, the $v_\mathrm{LOS}$ has a stripy pattern with the same orientation as the fibrils at chromospheric heights. The micro-turbulent velocity is on average about 0.5~\kms\ higher in the fibrillar pixels compared to their surroundings. 


The response function at the head of the bright fibrils where they are closer to the magnetic concentrations, peaks at larger optical depths. This agrees with the morphology of bright fibrils bending down towards network areas as for example illustrated in \citet{Wedemeyer2008}. However, our inversion results are given on an optical-depth scale, and we cannot draw firm conclusions about the morphology of the fibrils on a geometrical height scale.

Bright fibrils also appear in synthetic images in the line core of \CaIIIR, \CaIIK, \Halpha, and \MgIIK\ \citep{johan2019} computed from a 3D MHD simulation performed using the MURaM code \citep{rempelmuram,2018NatAs.tmp..173C}. The appearance of bright fibrils in this simulation is comparable to the observations, but the simulated bright fibrils are wider. We speculate that the difference in fibril width  is at least partially caused by the effective resolution of the simulation ($\sim 250$~km), which is larger than the resolution of the observations.

There are only very few bright fibrils in \CaIIK\ images computed from the simulation by \citet{2016A&A...585A...4C} performed with the Bifrost code, and none at all in \Halpha\ images \citep{jorrit15}. The difference in the number of fibrils and the density with which they are packed between the Bifrost and MURaM simulations 
could be caused by the difference in magnetic activity between this simulation (48~G average unsigned flux in the photosphere) and the simulation of \citet[][$\sim300$~G]{johan2019}. We note however that the average unsigned flux in the photosphere in our observations is  $\sim800$~G, which is much larger than either simulation. An additional reason for the difference in appearance of the fibrils in the simulations could be the larger separation between the opposite-polarity footpoints of the magnetic structures in the MURaM simulation ($\sim50$~Mm) compared to the Bifrost simulation ($\sim8$~Mm). 

In this study we mainly focused on a sample of bright fibrils in \CaIIK\ observations and studied their physical characteristics only. Their formation process is still an open question that needs to be answered in future studies.

\begin{acknowledgements}
SK and JL were supported through the CHROMATIC project (2016.0019) funded by the Knut and Alice Wallenberg foundation.

SD acknowledges support from Vinnova through grant MSCA 796805.

JdlCR is supported by grants from the Swedish Research Council (2015-03994) and the Swedish National Space Board (128/15).

This project has received funding from the European Research Council (ERC) under the European Union's Horizon 2020 research and innovation programme (SUNMAG, grant agreement 759548). SD and JdlCR are supported by a grant from the Swedish Civil Contingencies Agency (MSB).

The Swedish 1-m Solar Telescope is operated on the island of La Palma by the Institute for Solar Physics of Stockholm University in the Spanish Observatorio del Roque de los Muchachos of the Instituto de Astrof\'isica de Canarias.

The Institute for Solar Physics is supported by a grant for research infrastructures of national importance from the Swedish Research Council (registration number 2017-00625).

The computations were performed on resources provided by the Swedish National Infrastructure for Computing (SNIC) at the National Supercomputer Centre at Link\"oping University and the High Performance Computing Center North at Ume\aa\ University.

This study has been discussed within the activities of team 399 'Studying magnetic-field-regulated heating in the solar chromosphere' at the International Space Science Institute (ISSI) in Switzerland.

\end{acknowledgements}


\bibliographystyle{aa} 
\bibliography{paper}

\end{document}